\documentclass[11pt,oneside]{article}

\usepackage[round,colon,authoryear]{natbib} 

\usepackage{amsmath}
\usepackage{amsfonts}
\usepackage{amsthm}
\usepackage{hyperref}
\usepackage{calrsfs}
\DeclareMathAlphabet{\pazocal}{OMS}{zplm}{m}{n}
\usepackage[mathscr]{euscript}
\usepackage{amssymb}
\usepackage[shortlabels]{enumitem}
\usepackage{stmaryrd}
\usepackage{bm}

\usepackage{graphicx}
\graphicspath{ {./images/PTBnew/} }
\usepackage{caption}
\usepackage{float}
\usepackage{subfloat}
\usepackage{subfig}
\usepackage{multirow}
\usepackage[table,xcdraw]{xcolor}
\usepackage{accents}			

\usepackage{bbm}					
\numberwithin{equation}{section}
 
\newtheorem{thm}{Theorem}[section]

\newtheorem{prop}[thm]{Proposition}

\theoremstyle{definition}
\newtheorem{defn}[thm]{Definition}

\theoremstyle{remark}
\newtheorem{example}[thm]{Example}

\theoremstyle{remark}
\newtheorem{rem}[thm]{Remark}

\title{
Market-to-book ratio in Stochastic Portfolio Theory
}

\vspace{2mm}

\author{  
\textsc{Donghan Kim} 
\thanks{ 
Department of Mathematics, University of Michigan, Ann Arbor, MI 48109 (E-mail: {\it donghank@umich.edu}).
}
}

\begin{document}

\maketitle

\bigskip

\begin{abstract}
	We study market-to-book ratios of stocks in the context of Stochastic Portfolio Theory. Functionally generated portfolios that depend on auxiliary economic variables other than relative capitalizations~(``sizes'') are developed in two ways, together with their relative returns with respect to the market. This enables us to identify the value factor~(i.e., market-to-book ratio) in returns of such generated portfolios when the auxiliary variables are stocks' book values. Examples of portfolios, as well as their empirical results, are given, with the evidence that, in addition to size, the value factor does affect the performance of the portfolio.
\end{abstract}

\bigskip

\section{Introduction}
\label{sec: intro}
In the three-factor model of \citet{Fama:French1993} the traditional asset pricing model, formally known as Capital Asset Pricing Model~(CAPM), was expanded by including two additional risk factors, size and value, in addition to market risk. The resulting three-factor regression model explains over 90\% of the diversified portfolio returns, versus 70\% given by the CAPM. It also highlights the historic excess returns of small-cap stocks over big-cap stocks, and of value stocks (with high book-to-market ratios) over growth stocks (with low book-to-market ratios).

\smallskip

Following the work on the three-factor model, there has been additional research by the same authors regarding the value factor, i.e., the book-to-market ratio, in the direction of explaining the expected returns of stocks. The evolution of book-to-market ratio in terms of past changes in book equity and price, is shown to contain independent information about expected cashflows that can enhance estimates of expected returns \cite{Fama:French2008}. The authors also show evidence that a higher book-to-market ratio implies a higher expected stock return, along with two additional variables - expected profitability and expected investment - which are augmented to the five-factor model \cite{Fama:French2006}, \cite{Fama:French2015}. Similar evidence can be found in related works by other authors, for example, \cite{Rosenberg1985}, \cite{Chan1991}, \cite{Capaul1993}, \cite{Lakonishok1994}, and \cite{Cohen2003}.

\smallskip

Along a parallel development, the size factor of portfolio's return has been studied theoretically within a mathematical framework called Stochastic Portfolio Theory~(SPT), introduced by \citet{Fe}. SPT models the equity market using random processes, provides a method called functional generation of portfolios to construct a variety of portfolios from a function depending on individual companies' capitalizations, and analyzes portfolio behavior in a model-free, robust manner. SPT also explains under what structural conditions it becomes possible to outperform a capitalization-weighted benchmark index, and casts these conditions in very precise mathematical terms.

\smallskip

For example, the equal-weighted and so-called diversity-weighted portfolios, which favor smaller firms by according them the same~(equal-weighted), or more~(diversity-weighted), importance as large-cap companies, are generated from concave functions and are shown to outperform the capitalization-weighted index under suitable conditions. Then, the excess return rate of these portfolios over the benchmark index can be calculated from the mathematical framework. Moreover, SPT also handles rank-based portfolios, and enables comparisons of two capitalization-weighted portfolios: one consisting of a fixed number of big-cap stocks, and the other composed of small-cap stocks. By computing the relative return of the large-stock index with respect to the small-cap index, a term called leakage is observed, which measures the loss in relative return due to stocks that decline and are subsequently dropped from the `upper capitalization league' and replaced by others.

\smallskip

Though there is considerable evidence that the size of stocks does affect the portfolio return \cite{Fe, FK_survey, Banner_alpha}, SPT has not dealt with the value factor so far. We adopt in this paper the methodologies of SPT, in order to incorporate market-to-book ratios of stocks in the study of portfolio returns. More specifically, we first present a way of generating portfolios which depend on additional economic variables having paths of finite variation, besides each company's capitalization; we compute their relative returns with respect to the market~(capitalization-weighted index), and provide conditions under which such portfolios can outperform the market.

\smallskip

The additional variables can be any stochastic processes investors observe from the market; we even allow them to have discontinuous~(but right-continuous) paths, as far as they are of finite variation. In this case, we can construct portfolios in a similar manner under a mild condition on generating functions. Both of the portfolio weights and the relative returns with respect to the market are then stochastic processes with left-continuous paths. To the author's knowledge, functionally generated portfolios involving such discontinuities is new, in the context of SPT.

\smallskip

If we choose relative book values~(defined in \eqref{def : beta}) as the auxiliary process, then we can construct portfolios that depend on the market-to-book ratios of stocks. Many examples of such portfolios will be introduced, including the ones depending on the ranks of market-to-book ratios. We also provide empirical results that identify the value factor in portfolio returns.

\smallskip

\noindent
\textit{Preview} : Section~\ref{sec: setup} introduces a mathematical setup of a stock market, and defines trading strategies and portfolios. Section~\ref{sec: FGP} discusses how to generate portfolios depending on stocks' capitalizations and an additional process in two ways, additive and multiplicative, along with rank based portfolios. Section~\ref{sec: disconti} considers the case when the additional process is not necessarily continuous. Section~\ref{sec: examples} provides several examples of such portfolios depending on the book values, or the market-to-book ratios, and Sections~\ref{sec: PTB component} introduces a way to measure the market-to-book ratio component from portfolio return for general portfolios. Section~\ref{sec: empirical results} contains empirical results regarding the portfolios discussed in earlier sections.

\bigskip

\section{Setup}
\label{sec: setup}

\subsection{The stock market}

Let us suppose that a stock market is composed of $d$ stocks. We assume that trading is continuous in time, with no transaction costs or taxes, and that shares are infinitely divisible. We consider a vector $S = (S_1, \cdots, S_d)'$ of $d$ positive continuous semimartingales on a filtered probability space $(\Omega, \mathcal{F}, \mathbb{F}, \mathbb{P})$, which represent the prices of $d$ stocks in this market. Without loss of generality, we posit that each stock has always a single share outstanding, so that the price $S_i(t)$ of the $i$-th stock is equal to its capitalization at any time $t \ge 0$ for $i = 1, \cdots, d$.

\smallskip

From the capitalization vector process $S$, we define the vector process $\mu = (\mu_1, \cdots, \mu_d)'$ with components
\begin{equation}	\label{def : market weights}
\mu_i(t) := \frac{S_i(t)}{\sum_{j=1}^d S_j(t)}, \qquad i = 1, \cdots, d, \qquad t \ge 0.
\end{equation}
These are the individual companies' \textit{market weights}, and satisfy
\begin{equation*}
\sum_{j=1}^d \mu_j(t) \equiv 1, \qquad \forall~t \ge 0.
\end{equation*}
Note that each $\mu_i$ is also a positive continuous semimartingale, and represents the relative capitalization of the $i$-th stock with respect to the total capitalization of the market.

\medskip

\subsection{Trading strategies and portfolios}
We introduce the following notion of trading strategy.
\begin{defn} [Trading strategy and its relative wealth process]	\label{Def : investment}
	We call a $d$-dimensional vector of predictable process $\vartheta = (\vartheta_1, \cdots, \vartheta_d)'$ \textit{trading strategy}, if it is integrable with respect to the market weight vector $\mu$, and if the relative wealth process
	\begin{equation}		\label{def : wealth process}
	V^{\vartheta}(t) := \sum_{i=1}^d \vartheta_i(t)\mu_i(t) = \frac{\sum_{i=1}^d\vartheta_i(t)S_i(t)}{\sum_{i=1}^d S_i(t)}, \qquad \forall~t \ge 0
	\end{equation}
	corresponding to it, satisfies the `self-financing' identity
	\begin{equation}	\label{eq : self-financing}
	V^{\vartheta}(t) = V^{\vartheta}(0) + \int_0^t \sum_{i=1}^d \vartheta_i(u)d\mu_i(u), \qquad \forall~t \ge 0.
	\end{equation}
\end{defn}

\smallskip

The $i$-th component $\vartheta_i(t)$ of the trading strategy $\vartheta$ represents the number of shares held in the $i$-th stock at time $t \ge 0$, whereas the product $\vartheta_i(t)S_i(t)$ represents the amount of wealth invested in the $i$-th stock at time $t$. The last equation of \eqref{def : wealth process} follows from \eqref{def : market weights}, and $V^{\vartheta}(t)$ represents the ratio of the total wealth generated by the trading strategy $\vartheta$, to the total market capitalization at time $t \ge 0$. This interpretation justifies its name `relative wealth' with respect to the market, and this quantity $V^{\vartheta}(t)$ measures the extent to which the trading strategy $\vartheta$ outperforms (or underperforms) the market over time $t \ge 0$.

\medskip

\begin{defn} [Portfolio]	\label{Def : portfolio}
	For a trading strategy $\vartheta$, we call the vector $\pi^{\vartheta} = (\pi^{\vartheta}_1, \cdots, \pi^{\vartheta}_d)'$ with components
	\begin{equation}	\label{def : portfolio}
	\pi^{\vartheta}_i := \frac{\vartheta_iS_i}{\sum_{j=1}^d \vartheta_jS_j} 
	= \frac{\vartheta_i\mu_i}{\sum_{j=1}^d \vartheta_j\mu_j}, \qquad i = 1, \cdots, d,
	\end{equation}
	the \textit{portfolio} corresponding to the trading strategy $\vartheta$.
\end{defn}

\smallskip

The component $\pi^{\vartheta}_i(t)$ of the portfolio $\pi^{\vartheta}$ represents the proportion of wealth invested in the $i$-th stock at time $t \ge 0$, for each $i=1, \cdots, d$. We also denote by $V^{\pi} \equiv V^{\vartheta}$ the relative wealth process generated by the portfolio $\pi$ that corresponds to the trading strategy $\vartheta$.

\smallskip

One important portfolio is the one consisting of the market weights $\mu=(\mu_1, \cdots \mu_d)'$, defined in \eqref{def : market weights}, which we also call \textit{market portfolio}. This is accomplished by setting $\vartheta_i$ to a fixed constant, same for every $i$ in \eqref{def : portfolio}, i.e., by buying a fixed number of shares in each stock and holding on the these shares at all times. Then, the value of market portfolio is proportional to the total capitalization of the entire stock market. In this sense, the relative wealth process $V^{\vartheta}$ in \eqref{def : wealth process} of $\vartheta$ just represents the ratio of the value of $\vartheta$ with respect to the value of market portfolio $\mu$.

\bigskip

\section{Functional Generation of Portfolios}
\label{sec: FGP}

Functional portfolio generation, introduced by \citet{F_generating} more than 20 years ago, is a method of constructing trading strategies using a generating function that acts solely on the market weight vector $\mu$ of \eqref{def : market weights}. \citet{Karatzas:Ruf:2017} developed another approach for constructing trading strategies, which they call `additive generation' of portfolios, as opposed to Fernholz's `multiplicative generation'.

\smallskip

Moreover, the construction of these two types of trading strategies can be generalized by allowing generating functions to depend on an additional argument of finite variation other than the market weights; see \cite{Strong}, \cite{Schied:2016}, \cite{Ruf:Xie}, and \cite{Karatzas:Kim}. In this section, we summarize the ideas of constructing portfolios in terms of the market weight vector $\mu$, along with an additional auxiliary argument.

\medskip

\subsection{Generating function and its Gamma process}

Let us consider a function $G : [0, 1]^d \times \mathbb{R}^d \rightarrow \mathbb{R}$ which takes the vector process $\mu$ of market weights and an additional $d$-dimensional vector process $\Lambda$ of finite variation as its arguments. Each component $\Lambda_i$ of $\Lambda$ can be interpreted to model an observable quantity related to the $i$-th company, other than its relative capitalization $\mu_i$; for instance, though not exclusively, its book value. We assume in this section that each $\Lambda_i$ is a stochastic process with continuous path.

\smallskip

Assuming $G$ to be twice continuously differentiable with respect to the components of $\mu$ and continuously differentiable with respect to the components of $\Lambda$, we define a continuous, adapted process $\Gamma^{G}(\cdot)$, called the \textit{Gamma process} corresponding to the function $G$, as
\begin{equation}	\label{def : Gamma mu}
\Gamma^{G}(t) := G\big(\mu(0), \Lambda(0)\big) - G\big(\mu(t), \Lambda(t)\big) + \int_0^t \sum_{i=1}^d D_{\mu_i}G\big(\mu(s), \Lambda(s)\big)d\mu_i(s), \qquad t \ge 0.
\end{equation}
An application of It\^o's rule yields the alternative representation
\begin{align}	
\Gamma^{G}(t) = &-\frac{1}{2} \int_0^t \sum_{i=1}^d\sum_{j=1}^d D^2_{\mu_i, \mu_j} G\big(\mu(s), \Lambda(s)\big) d\langle \mu_i, \mu_j \rangle(s)			\label{eq : Gamma mu}
\\
& + \int_0^t \sum_{i=1}^d D_{\Lambda_i}G\big(\mu(s), \Lambda(s)\big) d\Lambda_i(s), \qquad t \ge 0,		\nonumber
\end{align}
which shows that the process $\Gamma^{G}(\cdot)$ is of finite-variation on compact time-intervals.

\smallskip

We show in the next subsection that the relative wealth process $V^{\vartheta}$ in \eqref{def : wealth process} of a trading strategy $\vartheta$, generated either additively or multiplicatively by a function $G$, can be expressed in terms of $G$ and of the Gamma process $\Gamma^G$; and that this finite-variation process $\Gamma^G$ is responsible for the long-run relative performance of the trading strategy with respect to the market.

\medskip

\subsection{Refined functional generation of portfolios}	\label{subsec : FGP}

In order to construct trading strategies which outperform the market, we may want to choose $G$ such that $\Gamma^G(\cdot)$ is nondecreasing. The first integral on the right-hand side of \eqref{eq : Gamma mu} will be nondecreasing in $t$, if $G$ is concave in $\mu$. However, it is quite hard to make the second integral nondecreasing, as the integrators $\Lambda_i$ are not generally monotone. 

\smallskip

To help deal with this issue, we consider canonical decompositions
\begin{equation}		\label{eq : Lambda}
\Lambda_i := \gamma_i - \xi_i, \qquad i = 1, \cdots, d
\end{equation}	
for the auxiliary processes $\Lambda_i(\cdot)$, where $\gamma_i(\cdot)$ and $\xi_i(\cdot)$ are the smallest nondecreasing, adapted processes with $\gamma_i(0) = \Lambda_i(0)$, for which \eqref{eq : Lambda} holds. We note that every $\xi_i$ is nonnegative with $\xi_i(0) = 0$ for $i = 1, \cdots, d$. We let the generating function $G$ depend on all three arguments $\mu, \gamma, \xi$, then we have the following new definition of \eqref{def : Gamma mu} and the representation of \eqref{eq : Gamma mu}
\begin{align}
\Gamma^G(t) := \, & \,G(\mu, \gamma, \xi)(0) -  G(\mu, \gamma, \xi)(t) + \int_0^t \sum_{i=1}^d D_{\mu_i}G(\mu, \gamma, \xi)(s) d\mu_i(s)			\nonumber
\\
= &- \frac{1}{2} \int_0^t \sum_{i, j = 1}^d D^2_{\mu_i, \mu_j} G(\mu, \gamma, \xi)(s) d\langle \mu_i, \mu_j \rangle (s)			\label{eq : Gamma2 mu}
\\
& - \int_0^t \sum_{i=1}^d D_{\gamma_i}G(\mu, \gamma, \xi)(s) d\gamma_i(s)
- \int_0^t \sum_{i=1}^d D_{\xi_i}G(\mu, \gamma, \xi)(s) d\xi_i(s), \qquad t \ge 0.		\nonumber
\end{align}
We are assuming, tacitly, that the function $G$ and all its indicated derivatives are continuous. It is now clear that the Gamma process $\Gamma^G(\cdot)$ is nondecreasing, if $G$ is concave with respect to $\mu$, and if
\begin{equation}	\label{Eq : two inequalities mu}
D_{\gamma_i}G(\mu, \gamma, \xi) \le 0, \qquad  D_{\xi_i}G(\mu, \gamma, \xi) \le 0, \qquad i = 1, \cdots, d.
\end{equation}
Examples of portfolios satisfying the condition \eqref{Eq : two inequalities mu} will be given in Subsection~\ref{subsec: Examples of AGP}.

\medskip

For a given function $G$, we construct now two different (additively and multiplicatively generated) types of trading strategies, as well as their corresponding portfolios, under appropriate conditions, using the vector of market weights $\mu$, and two vectors $\gamma$, $\xi$, composed of nondecreasing components.

\smallskip

In the following, we denote by $supp(X)$ the support of a $d$-dimensional stochastic process $X$, i.e., the smallest closed set $C \subset \mathbb{R}^d$ satisfying $\mathbb{P}(X(t) \in C, ~\forall \, t \ge 0) = 1$. We also denote by $C^{2, 1, 1}$~($C^{2, 1}$) the collection of functions $G : \mathbb{R}^{3d} \rightarrow \mathbb{R}$~($G : \mathbb{R}^{2d} \rightarrow \mathbb{R}$) such that $G$ is twice continuously differentiable with respect to the first $d$ components and continuously differentiable with respect to the last $2d$~($d$) components, respectively.

\smallskip

For the proofs of Propositions~\ref{prop : additive generation mu} and \ref{prop : multiplicative generation mu} below, we defer to Theorems~\ref{thm : additive generation disconti} and \ref{thm : multiplicative generation disconti}, or we refer readers to Propositions~4.3, 4.8 of \cite{Karatzas:Ruf:2017}, or Propositions~3.6, 3.9 of \cite{Karatzas:Kim}.

\smallskip

\begin{prop} [\textit{Additive Functional Generation}]	\label{prop : additive generation mu}
	For a given function $G:supp(\mu) \times supp(\gamma) \times supp(\xi) \rightarrow \mathbb{R}$ in $C^{2, 1, 1}$, we define two vector processes $\vartheta = (\vartheta_1, \cdots, \vartheta_d)'$ and $\varphi = (\varphi_1, \cdots, \varphi_d)'$ with components
	\begin{align}	
	\vartheta_i(t) &:= D_{\mu_i}G(\mu, \gamma, \xi)(t), 	\label{def : vartheta mu}
	\\	
	\varphi_i(t) &:= \vartheta_i(t) - Q^{\vartheta, \mu}(t)-C^{G}(0), 			\label{def : varphi mu}
	\end{align}
	for $i = 1, \cdots, d$, and $t \ge 0$. Here, the process
	\begin{equation}	\label{def : defect of sf mu}
	Q^{\vartheta, \mu}(t) := \sum_{i=1}^d \vartheta_i(t)\mu_i(t)-\sum_{i=1}^d \vartheta_i(0)\mu_i(0)-\int_0^t\sum_{i=1}^d \vartheta_i(s)d\mu_i(s), \qquad t \ge 0
	\end{equation}
	is called the `defect of self-financibility', and the real constant $C^{G}(0)$ is the value at time $t=0$ of the `defect of balance' process
	\begin{equation}	\label{def : defect of b mu}
	C^{G}(\cdot)
	:= \sum_{i=1}^d \mu_i(\cdot)\vartheta_i(\cdot)-G(\mu, \gamma, \xi)(\cdot) 
	= \sum_{i=1}^d \mu_i(\cdot)D_{\mu_i}G(\mu, \gamma, \xi)(\cdot)-G(\mu, \gamma, \xi)(\cdot).
	\end{equation}
	
	Then the vector process $\varphi$ of \eqref{def : varphi mu} is a trading strategy, which can be represented in the form
	\begin{equation}	\label{eq : varphi mu}
	\varphi_i(t) = \Gamma^{G}(t)+\vartheta_i(t)-C^{G}(t), \qquad t \ge 0,
	\end{equation}
	for $i=1, \cdots, d$, and generates the relative wealth process $V^{\varphi}$, given as
	\begin{equation}	\label{eq : AG wealth mu}
	V^{\varphi}(t) = G(\mu, \gamma, \xi)(t) + \Gamma^{G}(t), \qquad t \ge 0.
	\end{equation}
\end{prop}

\smallskip

\begin{prop} [\textit{Multiplicative Functional Generation}]	\label{prop : multiplicative generation mu}
	For a given function $G:supp(\mu) \times supp(\gamma) \times supp(\xi) \rightarrow (0, \infty)$ in $C^{2, 1, 1}$ such that the process $1/G(\mu, \gamma, \xi)$ is locally bounded, we recall the vector process $\vartheta = (\vartheta_1, \cdots, \vartheta_d)'$ of \eqref{def : vartheta mu} and define two vector processes $\eta = (\eta_1, \cdots, \eta_d)'$ and $\psi = (\psi_1, \cdots, \psi_d)'$ with components
	\begin{align}
	\eta_i(t) &:= \vartheta_i(t) \cdot \exp\bigg(\int_0^t\frac{d\Gamma^{G}(s)}{G(\mu, \gamma, \xi)(s)}\bigg),			\label{def : eta mu}
	\\
	\psi_i(t) &:= \eta_i(t) - Q^{\eta, \mu}(t)-C^{G}(0),		\nonumber
	\end{align}
	for $i = 1, \cdots, d,$ and $t \ge 0$, where $Q^{\eta, \mu}$ and $C^{G}(0)$ are defined as in \eqref{def : defect of sf mu} and \eqref{def : defect of b mu}. Then $\psi$ is a trading strategy, and can be represented in the form
	\begin{equation}	\label{eq : psi mu}
	\psi_i(t) = G(\mu, \gamma, \xi)(t) \cdot \exp\bigg(\int_0^t\frac{d\Gamma^{G}(s)}{G(\mu, \gamma, \xi)(s)}\bigg) + \eta_i(t) - \sum_{j=1}^d \mu_j(t) \eta_j(t), \qquad t \ge 0,
	\end{equation}
	for $i=1, \cdots, d$. The relative wealth process $V^{\psi}$ it generates, is given as
	\begin{equation}	\label{eq : MG wealth mu}
	V^{\psi}(t) =  G(\mu, \gamma, \xi)(t) \cdot \exp\bigg(\int_0^t\frac{d\Gamma^{G}(s)}{G(\mu, \gamma, \xi)(s)}\bigg), \qquad t \ge 0.
	\end{equation}
\end{prop}

\smallskip

\begin{defn} [Functionally generated trading strategies] \label{Def : FGTS mu}
	The trading strategies $\varphi$ of Proposition~\ref{prop : additive generation mu}, and $\psi$ of Proposition~\ref{prop : multiplicative generation mu}, are called \textit{additively generated trading strategy} and \textit{multiplicatively generated trading strategy} from $G$, respectively. This terminology is justified by the expressions in \eqref{eq : AG wealth mu}, \eqref{eq : MG wealth mu}.
\end{defn}

\medskip

\noindent \underline{Important Remark:} When the generating function $G$ satisfies the identity
\begin{equation}	\label{def : balanced mu}
G(\mu, \gamma, \xi) = \sum_{i=1}^d \mu_i \cdot D_{\mu_i}G(\mu, \gamma, \xi),
\end{equation}
we say that it is \textit{balanced}. We see that, for balanced generating functions $G$, the ``defect-of-balance'' process $C^{G}(\cdot)$ is identically equal to zero in \eqref{def : defect of b mu}, and the representations of the trading strategies $\varphi$ in \eqref{eq : varphi mu} and $\psi$ in \eqref{eq : psi mu} are simplified significantly as
\begin{align}
\varphi_i(t) &= \Gamma^{G}(t)+\vartheta_i(t), && t \ge 0,		\label{eq : balanced varphi mu}
\\
\psi_i(t) &= \eta_i(t), && t \ge 0,								\label{eq : balanced psi mu}
\end{align}
in the notation of \eqref{eq : Gamma2 mu}, \eqref{def : vartheta mu}, and \eqref{def : eta mu}. We will see later in Section~\ref{sec: disconti} that this balance condition plays an important role in the construction of trading strategies, when the components of the auxiliary process $\Lambda$ are discontinuous.

\bigskip

\subsection{Strong relative arbitrage}

A judicious choice of generating function produces a trading strategy with the potential of outperforming the market portfolio $\mu$ of \eqref{def : market weights}, under appropriate conditions. We introduce in this subsection the notion of strong relative arbitrage, and sufficient conditions leading to it.

\smallskip

\begin{defn} [Strong relative arbitrage]
	We say that a given trading strategy $\vartheta$ is \textit{strong relative arbitrage with respect to the market over the time horizon $[0, T]$}, if we have
	\begin{equation}
	V^{\vartheta}(t) \ge 0 , \qquad \forall~t \in [0, T],
	\end{equation}
	and
	\begin{equation}
	\mathbb{P}[V^{\vartheta}(T) > V^{\vartheta}(0)] = 1.
	\end{equation}
\end{defn}

\smallskip

Since the additively generated trading strategies lead to simpler sufficient conditions for strong relative arbitrage than the multiplicatively generated ones, we present here only sufficient condition for additive portfolio generation, based on Theorem~5.1 of \cite{Karatzas:Ruf:2017}. We refer interested readers to Theorem~5.2 of \cite{Karatzas:Ruf:2017} and Subsection 4.2 of \cite{Karatzas:Kim}, for sufficient conditions of multiplicatively generated trading strategies leading to relative arbitrage. 

\smallskip

\begin{prop} [Additively generated strong relative arbitrage with respect to the market]	\label{prop : AGSRA mu}
	Fix a nonnegative function $G : supp(\mu) \times supp(\gamma) \times supp(\xi) \rightarrow [0, \infty)$ in $C^{2, 1, 1}$ such that the Gamma process $\Gamma^{G}(\cdot)$ is nondecreasing. For some real number $T_* > 0$, suppose that we have
	\begin{equation}	\label{con : AGSRA mu}
	\mathbb{P}[ \Gamma^{G}(T_*) > G(\mu, \gamma, \xi)(0) ] = 1.
	\end{equation}
	Then the trading strategy $\varphi$ additively generated from $G$, is a strong relative arbitrage with respect to the market portfolio $\mu$ over every time interval $[0, t]$ with $t \ge T_*$. 
\end{prop}

\medskip

\begin{rem}
	In order to obtain relative wealth processes starting from $1$ at the initial time, i.e., $V^{\varphi}(0) = V^{\psi}(0) = 1$ in \eqref{eq : AG wealth mu} and \eqref{eq : MG wealth mu}, we shall often normalize the generation function $G$ to satisfy $G(\mu, \gamma, \xi)(0) = 1$ by replacing $G(\mu, \gamma, \xi)(t)$ by
	\begin{equation*}
	\begin{cases}
	G(\mu, \gamma, \xi)(t)/G(\mu, \gamma, \xi)(0), \qquad &\text{if } G(\mu, \gamma, \xi)(0) > 0,
	\\
	G(\mu, \gamma, \xi)(t)+1, \qquad &\text{if } G(\mu, \gamma, \xi)(0) = 0.
	\end{cases}
	\end{equation*}	  
\end{rem}

\medskip

\subsection{Rank based generation of portfolios}	\label{subsec: ranked}

Recall the market weight vector $\mu$ in \eqref{def : market weights} and the auxiliary vector $\Lambda$ of finite variation. We construct trading strategies depending on the ranks of the components of the vector process $f(\mu, \Lambda)$, where $f : [0, 1]^d \times \mathbb{R}^d \rightarrow \mathbb{R}^d$ is a sufficiently smooth function. In order to simplify notation, we do not consider the decomposition of $\Lambda$ as in \eqref{eq : Lambda}, and let $f$ have two arguments $\mu$ and $\Lambda$. 

\smallskip

We introduce the $k$-th \textit{rank process} of any given $d$-dimensional vector process $\nu = (\nu_1, \cdots, \nu_d)'$, namely
\begin{equation*}
\nu_{[k]}(t) := \max_{1 \le i_1 < \cdots < i_k \le d} \min \big\{\nu_{i_1}(t), \cdots, \nu_{i_k}(t)\big\}, \qquad t \ge 0
\end{equation*}
for $k = 1, \cdots, d$. More specifically, we rank the components of the vector $\nu$ in descending order
\begin{equation*}
\nu_{[1]}(t) \ge \nu_{[2]}(t) \ge \cdots \ge \nu_{[d]}(t), \qquad t \ge 0,
\end{equation*}
with the lexicographic rule for breaking ties that always assigns a higher rank~(a smaller $[k]$) to the smaller index $i$. We also consider the random permutation process $r_t$ of $\{1, \cdots, d\}$ such that
\begin{equation}	\label{def : random permutation r}
\begin{split}
\nu_{r_t(k)}(t) &= \nu_{[k]}(t),
\\
r_t(k) < r_t(k+1) &\qquad \text{if} \qquad \nu_{[k]}(t) = \nu_{[k+1]}(t)
\end{split}
\end{equation}
hold for every $t \ge 0$ and $k = 1, \cdots, d$.

\smallskip

We shall use the notation $\pazocal{R}(\nu) := (\nu_{[1]}, \cdots, \nu_{[d]})'$, the vector arranged in order of descending rank. In particular, when $\nu \equiv f(\mu, \Lambda)$, that is, $\nu_i(\cdot) \equiv f_i(\mu(\cdot), \Lambda(\cdot))$ for $i = 1, \cdots, d$, we have $\pazocal{R}\big(f(\mu, \Lambda)\big) = \big(f_{[1]}(\mu, \Lambda), \cdots, f_{[d]}(\mu, \Lambda)\big)'$, satisfying $f_{[1]}(\mu, \Lambda) \ge \cdots \ge f_{[d]}(\mu, \Lambda)$.

\smallskip

\begin{example}	[Ranked market-to-book ratios]	\label{ex : ranked market-to-book ratio}
	$\nu_i$ is equal to the market-to-book ratio $\rho_i$, defined in \eqref{def : rho} later, if $\Lambda$ is the relative book value vector $\beta$ of \eqref{def : beta} and $f_i(\mu, \Lambda) = \mu_i/\Lambda_i = \rho_i$ for $i = 1, \cdots, d$. In this case, $r_t(k)$ is the index, or name, of the stock whose market-to-book ratio occupies the $k$-th rank at time $t$.
\end{example}

\medskip

We next denote $L^Y(t)$ the (semimartingale) local time at $0$ for a semimartingale $Y$:
\begin{equation*}
L^Y(t) := \frac{1}{2}\Big( \vert Y(t) \vert - \vert Y(0) \vert - \int_0^t \text{sgn}\big(Y(s)\big) dY(s) \Big),
\end{equation*}
where $\text{sgn}(x) := 2\cdot I_{(0, \infty)}(x)-1$. To simplify computations involving local time terms throughout this section, we need the following definition and assumption.
\begin{defn}
	A vector process $X=(X_1, \cdots, X_d)'$ is \textit{pathwise mutually nondegenerate} if
	\begin{enumerate}[(i)]
		\item for all $i \ne j$,  $\{t: X_i(t) = X_j(t)\}$ has Lebesgue measure zero, a.s.;
		\item for all $i < j < k$,  $\{t: X_i(t) = X_j(t) = X_k(t)\} = \emptyset$, a.s.
	\end{enumerate}
\end{defn}

\smallskip

\noindent \textbf{Assumption:} For a given function $f : [0, 1]^d \times \mathbb{R}^d \rightarrow \mathbb{R}^d$, the vector process $f(\mu, \Lambda)$ is pathwise mutually nondegenerate.

\medskip

Consider a twice continuously differentiable function $G$ which takes the ranked vector process $\pazocal{R}\big(f(\mu, \Lambda)\big)$ as an argument. In order to simplify the notation, we write $\pazocal{R}\big(f(\mu, \Lambda)\big) \equiv \bm{\nu}$, and apply It\^o's rule to obtain
\begin{align}	
G\Big( \pazocal{R}\big(f(\mu, \Lambda)\big) (t) \Big) 
&= G\big( \bm{\nu}(0)\big)			
+ \int_0^t \sum_{k=1}^d D_{k}G\big(\bm{\nu}(s)\big) df_{[k]}(\mu, \Lambda)(s)			\label{eq : ranked Ito}
\\
& \quad + \frac{1}{2} \int_0^t \sum_{k=1}^d \sum_{\ell=1}^d D_{k, \ell}^2 G\big(\bm{\nu}(s)\big) d\langle f_{[k]}(\mu, \Lambda), f_{[\ell]}(\mu, \Lambda) \rangle(s).		\nonumber
\end{align}
Corollary~2.6 of \cite{Banner:Ghomrasni}, along with the assumption about pathwise mutual nondegeneracy, gives the dynamics of ranked components
\begin{align}
df_{[k]}(\mu, \Lambda)(s)
= \sum_{j=1}^d I_{\{r_s(k)=j\}} df_j(\mu, \Lambda)(s) 
&+ \frac{1}{2}dL^{f_{[k]}(\mu, \Lambda)-f_{[k+1]}(\mu, \Lambda)}(s) 	\label{eq : ranked mu integral}
\\
&- \frac{1}{2}dL^{f_{[k-1]}(\mu, \Lambda)-f_{[k]}(\mu, \Lambda)}(s),	\nonumber
\end{align}
for $k = 1, \cdots, d$, with the convention $f_{[0]}(\mu, \Lambda) \equiv f_{[d+1]}(\mu, \Lambda) \equiv 0$. Another application of It\^o's rule yields
\begin{align}
df_j(\mu, \Lambda)&(t) 
= \sum_{i=1}^d D_{\mu_i}f_j(\mu, \Lambda)(t) d\mu_i(t)
\\
&+ \frac{1}{2}\sum_{m=1}^d \sum_{n=1}^d D^2_{\mu_m, \mu_n} f_j(\mu, \Lambda)(t) d \langle \mu_m, \mu_n \rangle(t)
+ \sum_{m=1}^d D_{\Lambda_m}f_j(\mu, \Lambda)(t) d\Lambda_m(t),	\nonumber
\end{align}
provided that $f_j \in C^{2, 1}$ for $j = 1, \cdots, d$.

\smallskip

We consider the following new definitions
\begin{equation}	\label{def : ranked vartheta mu}
\vartheta_i(t) := \sum_{j=1}^d \sum_{k=1}^d D_{k}G\big(\bm{\nu}(t)\big) I_{\{r_t(k)=j\}} D_{\mu_i}f_j(\mu, \Lambda)(t), \qquad t \ge 0,
\end{equation}
\begin{equation}	\label{def : ranked Gamma mu}
\Gamma^G(t) := G\big(\bm{\nu}(0)\big) - G\big(\bm{\nu}(t)\big) + \int_0^t \sum_{i=1}^d \vartheta_i(s) d\mu_i(s), \qquad t \ge 0,
\end{equation}
analogues of \eqref{def : vartheta mu}, \eqref{eq : Gamma2 mu}. Then, the Gamma function $\Gamma^G$ can also be cast with the help of \eqref{eq : ranked Ito}-\eqref{def : ranked Gamma mu}, as
\begin{align}
\Gamma^G(t) = &-\int_0^t \sum_{k, j, m = 1}^d D_{k}G\big(\bm{\nu}(s)\big) I_{\{r_s(k)=j\}} D_{\Lambda_m} f_j(\mu, \Lambda)(s) d\Lambda_m(s) 		\nonumber
\\
&-\frac{1}{2} \int_0^t \sum_{j, k, m, n = 1}^d D_k G\big(\bm{\nu}(s)\big) I_{\{r_s(k)=j\}} D^2_{\mu_m, \mu_n} f_j(\mu, \Lambda)(s)  d\langle \mu_m, \mu_n \rangle(s)	\nonumber
\\
&-\frac{1}{2} \int_0^t \sum_{k, \ell, i, j, m, n = 1}^d D^2_{k, \ell}  G\big(\bm{\nu}(s)\big) I_{\{r_s(k)=i\}} I_{\{r_s(\ell)=j\}} D_{\mu_m} f_i(\mu, \Lambda)(s) D_{\mu_n} f_j(\mu, \Lambda)(s) d\langle \mu_m, \mu_n \rangle(s)	\nonumber
\\
&-\frac{1}{2}\int_0^t \sum_{k=1}^d D_{k}G\big(\bm{\nu}(s)\big) dL^{f_{[k]}(\mu, \Lambda)-f_{[k+1]}(\mu, \Lambda)}(s)	\nonumber
\\
&+\frac{1}{2}\int_0^t \sum_{k=1}^d D_{k}G\big(\bm{\nu}(s)\big) dL^{f_{[k-1]}(\mu, \Lambda)-f_{[k]}(\mu, \Lambda)}(s).	\label{eq : ranked Gamma mu}
\end{align}
This shows that the Gamma process of \eqref{def : ranked Gamma mu} is again of finite variation. 

\smallskip

Without the assumption about pathwise mutual nondegeneracy, we obtain a similar expression of the Gamma process, where the last two terms of \eqref{eq : ranked mu integral} and \eqref{eq : ranked Gamma mu} then involve higher orders of collision local time terms $L^{f_{[k]}(\mu, \Lambda)-f_{[\ell]}(\mu, \Lambda)}$ for $|k-\ell| \ge 2$.

\medskip

We construct now trading strategies from the given ranked vector process $\bm{\nu} \equiv \pazocal{R}\big(f(\mu, \Lambda)\big)$ in the same manner as in Subsection~\ref{subsec : FGP}.

\begin{prop} [\textit{Additive Functional Generation}]	\label{prop : ranked additive generation}
	For a twice continuously differentiable function $G:supp(\bm{\nu}) \rightarrow \mathbb{R}$, we define two vector processes $\vartheta = (\vartheta_1, \cdots, \vartheta_d)'$ as in \eqref{def : ranked vartheta mu} and $\varphi = (\varphi_1, \cdots, \varphi_d)'$ with components
	\begin{equation}
	\varphi_i(t) := \vartheta_i(t) - Q^{\vartheta, \mu}(t)-C^{\vartheta}(0), \qquad t \ge 0,
	\end{equation}
	for $i = 1, \cdots, d$. Here, we recall the defect of self-financibility $Q^{\vartheta, \mu}$ in \eqref{def : defect of sf mu}, and the defect of balance
	\begin{equation}	\label{def : ranked defect of b}
	C^{\vartheta}(t) := \sum_{i=1}^d \mu_i(t)\vartheta_i(t)-G\big(\bm{\nu}(t)\big), \qquad t \ge 0.
	\end{equation}
	
	Then $\varphi$ is a trading strategy, and can be represented in the form
	\begin{equation}	\label{eq : ranked varphi}
	\varphi_i(t) = \Gamma^{G}(t)+\vartheta_i(t)- C^{\vartheta}(t), \qquad t \ge 0,
	\end{equation}
	for $i=1, \cdots, d$, with $\Gamma^G$ defined in \eqref{def : ranked Gamma mu}. Its relative wealth process $V^{\varphi}$ is given as
	\begin{equation}	\label{eq : AG wealth s}
	V^{\varphi}(t) = G\big(\bm{\nu}(t)\big) + \Gamma^{G}(t), \qquad t \ge 0.
	\end{equation}
\end{prop}

\medskip

\begin{prop} [\textit{Multiplicative Functional Generation}]	\label{prop : ranked multiplicative generation}
	For a twice continuously differentiable function $G : supp(\bm{\nu}) \rightarrow (0, \infty)$ such that the process $1/G(\bm{\nu})$ is locally bounded, we recall the vector process $\vartheta = (\vartheta_1, \cdots, \vartheta_d)'$ from \eqref{def : ranked vartheta mu}, and define two vector processes $\eta = (\eta_1, \cdots, \eta_d)'$ and $\psi = (\psi_1, \cdots, \psi_d)'$ with components
	\begin{equation}		\label{def : ranked eta}
	\eta_i(t) := \vartheta_i(t) \cdot \exp\bigg(\int_0^t\frac{d\Gamma^{G}(s)}{G\big(\bm{\nu}(s)\big)}\bigg), \qquad t \ge 0,
	\end{equation}
	\begin{equation}
	\psi_i(t) := \eta_i(t) - Q^{\eta, \mu}(t)-C^{\eta}(0), \qquad t \ge 0,
	\end{equation}
	respectively. Here $Q^{\eta, \mu}$ and $C^{\eta}(0)$ are given as in \eqref{def : defect of sf mu}, \eqref{def : ranked defect of b}, and we recall the process $\Gamma^G(\cdot)$ from \eqref{def : ranked Gamma mu}. Then $\psi$ is a trading strategy, and can be represented in the form
	\begin{equation}	\label{eq : psi s}
	\psi_i(t) = G\big(\bm{\nu}(t)\big) \cdot \exp\bigg(\int_0^t\frac{d\Gamma^{G}(s)}{G\big(\bm{\nu}(s)\big)}\bigg) + \eta_i(t) - \sum_{j=1}^d \mu_j(t) \eta_j(t), \qquad t \ge 0,
	\end{equation}
	for $i=1, \cdots, d$. Its relative wealth process $V^{\psi}$ is given as
	\begin{equation}	\label{eq : MG wealth s}
	V^{\psi}(t) =  G\big(\bm{\nu}(t)\big) \cdot \exp\bigg(\int_0^t\frac{d\Gamma^{G}(s)}{G\big(\bm{\nu}(s)\big)}\bigg), \qquad t \ge 0.
	\end{equation}
\end{prop}

\bigskip

\section{Discontinuous auxiliary process}	\label{sec: disconti}

If some accounting information that affects the stock prices of the companies is updated periodically, say annually or quarterly, then it is more natural to model the components of the auxiliary process $\Lambda_1, \cdots, \Lambda_d$ as discontinuous processes. In this section, we allow each $\Lambda_i$ to be discontinuous, but assume it to be an RCLL process, i.e., right continuous with left limits. This accommodates jumps that occur whenever new accounting information is available.

\smallskip

Each company's relative capitalization $\mu_i$ of \eqref{def : market weights} remains a continuous process, whereas $\Lambda_i$, thus $\gamma_i$ and $\xi_i$ in the decomposition \eqref{eq : Lambda} are RCLL for $i = 1, \cdots d$. Here and in what follows, $\alpha^c(\cdot)$ and $\Delta \alpha(t) := \alpha(t+)-\alpha(t-)$ denote the continuous part and a jump at time $t$ of a stochastic process $\alpha$, respectively. In particular, $\Delta \alpha(t) = \alpha(t)-\alpha(t-)$ if $\alpha$ is RCLL~(c\`adl\`ag), whereas $\Delta \alpha(t) = \alpha(t+)-\alpha(t)$ if $\alpha$ is LCRL~(c\`agl\`ad). At $t = 0$, we have $\alpha(0-) = \alpha(0)$.

\smallskip

In order to handle jumps, we modify slightly the self-financing identity \eqref{eq : self-financing} in Definition~\ref{Def : investment}, as follows:	
\begin{equation}	\label{eq : self-financing disconti}
V^{\vartheta}(t) 
= V^{\vartheta}(0) + \int_0^t \sum_{i=1}^d \vartheta_i(s)d\mu_i(s) + \sum_{0 \le s < t} \sum_{i=1}^d \Delta \vartheta_i(s)\mu_i(s), \quad \forall ~ t \ge 0.
\end{equation}
Thus, the relative wealth process $V^{\vartheta}(\cdot) := \sum_{i=1}^d \vartheta_i(t)\mu_i(t)$ encompasses now possible jumps of the trading strategy $\vartheta$. We then define portfolio $\pi^{\vartheta}$ corresponding to $\vartheta$ as before in Definition~\ref{Def : portfolio}.

\smallskip

In order to construct functionally-generated portfolios, we apply It\^o's rule for jump processes to a function $G$ in $C^{2, 1, 1}$ to obtain
\begin{align}
& \quad G(\mu, \gamma, \xi)(t) 
= G(\mu, \gamma, \xi)(0) + \sum_{0 \le s \le t} \Delta G(\mu, \gamma, \xi)(s)	\label{eq : G1 disconti}
\\
& + \int_0^t \sum_{i=1}^d D_{\mu_i} G(\mu, \gamma, \xi)(s-) d\mu_i(s) 
+ \frac{1}{2} \int_0^t \sum_{i, j = 1}^d D_{\mu_i, \mu_j}^2 G(\mu, \gamma, \xi)(s-) d\langle \mu_i, \mu_j \rangle (s) 	\nonumber
\\
&+ \int_0^t \sum_{i=1}^d D_{\gamma_i}G(\mu, \gamma, \xi)(s-) d\gamma^c_i(s)	+ \int_0^t \sum_{i=1}^d D_{\xi_i}G(\mu, \gamma, \xi)(s-) d\xi^c_i(s),	\nonumber
\end{align}
where $\Delta G(\mu, \gamma, \xi)(s) := G(\mu, \gamma, \xi)(s)-G(\mu, \gamma, \xi)(s-)$. We define the Gamma process corresponding to the function $G$ in a similar manner, namely,
\begin{equation}	\label{def : Gamma1 disconti}
\Gamma^{G}(t) := G(\mu, \gamma, \xi)(0) - G(\mu, \gamma, \xi)(t) 
+ \int_0^t \sum_{i=1}^d D_{\mu_i}G(\mu, \gamma, \xi)(s-) d\mu_i(s),
\end{equation}
along with the alternative representation
\begin{equation*}
\Gamma^{G}(t) = \Gamma^{G, c}(t) - \sum_{0 \le s \le t} \Delta G(\mu, \gamma, \xi)(s),
\end{equation*}
where
\begin{align}
\Gamma^{G, c}(t) := & -\frac{1}{2} \int_0^t \sum_{i, j=1}^d D^2_{\mu_i, \mu_j}G(\mu, \gamma, \xi)(s-) d\langle \mu_i, \mu_j \rangle(s)	\label{eq : Gamma conti}
\\
&- \int_0^t \sum_{i=1}^d D_{\gamma_i}G(\mu, \gamma, \xi)(s-) d\gamma^c_i(s)
- \int_0^t \sum_{i=1}^d D_{\xi_i}G(\mu, \gamma, \xi)(s-) d\xi^c_i(s).	\nonumber
\end{align}
This shows that $\Gamma^{G}(\cdot)$ is of finite-variation.

\smallskip

We show in the following that the balance condition \eqref{def : balanced mu} is essential for handling the jumps when generating trading strategies.

\begin{thm} [\textit{Additive Functional Generation}]	\label{thm : additive generation disconti}
	For a function $G:supp(\mu) \times supp(\gamma) \times supp(\xi) \rightarrow \mathbb{R}$ in $C^{2, 1, 1}$ satisfying the balance condition \eqref{def : balanced mu}, we introduce two vector processes $\vartheta = (\vartheta_1, \cdots, \vartheta_d)'$ and $\varphi = (\varphi_1, \cdots, \varphi_d)'$ with components
	\begin{equation}	\label{def : vartheta disconti}
	\vartheta_i(t) := D_{\mu_i}G(\mu, \gamma, \xi)(t-), \qquad t \ge 0,
	\end{equation}
	\begin{equation}	\label{def : varphi disconti}
	\varphi_i(t) := \vartheta_i(t) - Q^{\vartheta, \mu}(t), \qquad t \ge 0,
	\end{equation}
	for $i = 1, \cdots, d$. Here, the defect of self-financibility is defined as
	\begin{align}
	Q^{\vartheta, \mu}(t)
	:= \sum_{i=1}^d \vartheta_i(t)\mu_i(t)-\sum_{i=1}^d \vartheta_i(0)\mu_i(0)
	&-\int_0^t\sum_{i=1}^d \vartheta_i(s)d\mu_i(s)	\label{def : defect of sf disconti}
	\\
	&- \sum_{0 \le s < t} \sum_{i=1}^d \Delta \vartheta_i(s)\mu_i(s), \qquad t \ge 0.			\nonumber
	\end{align}
	
	Then the vector process $\varphi$ in \eqref{def : varphi disconti} is a trading strategy, and can be represented in the form
	\begin{equation}	\label{eq : varphi disconti}
	\varphi_i(t) = \vartheta_i(t) + \Gamma^{G, c}(t), \qquad t \ge 0,
	\end{equation}
	for $i=1, \cdots, d$; whereas, the relative wealth process $V^{\varphi}$ it generates, is given as
	\begin{equation}	\label{eq : AG wealth disconti}
	V^{\varphi}(t) = G(\mu, \gamma, \xi)(t-) + \Gamma^{G, c}(t), \qquad t \ge 0.
	\end{equation}
\end{thm}

\begin{proof}
	We first note that $\vartheta_i$ defined in \eqref{def : varphi disconti} is left-continuous with $\vartheta_i(t+) = D_{\mu_i}G(\mu, \gamma, \xi)(t)$. The assertion that $\varphi$ is a trading strategy follows from Proposition 4.3 of \cite{Karatzas:Ruf:2017}. Thanks to the balance condition, we have the identities
	\begin{equation}	\label{eq: value G}
	\sum_{i=1}^d \vartheta_i(t)\mu_i(t) = \sum_{i=1}^d D_{\mu_i}G(\mu, \gamma, \xi)(t-) \mu_i(t) = G(\mu, \gamma, \xi)(t-), \qquad t \ge 0,
	\end{equation}
	as well as
	\begin{equation}	\label{eq: jump value G}
	\sum_{i=1}^d \Delta \vartheta_i(s)\mu_i(s)
	= \sum_{i=1}^d \Delta D_{\mu_i} G(\mu, \gamma, \xi)(s)\mu_i(s)
	= \Delta G\big(\rho(s)\big), \qquad s \ge 0.
	\end{equation}
	
	Substituting the definition \eqref{def : defect of sf disconti} into \eqref{def : varphi disconti}, along with \eqref{eq: value G} and \eqref{eq: jump value G}, we obtain:
	\begin{align*}
	\varphi_i(t) 
	&= \vartheta_i(t) - \sum_{j=1}^d \vartheta_j(t) \mu_j(t) + \sum_{j=1}^d \vartheta_j(0)\mu_j(0) + \int_0^t \sum_{j=1}^d \vartheta_j(s)d\mu_j(s) + \sum_{0 \le s < t}\sum_{j=1}^d \Delta \vartheta_j(s)\mu_j(s) \\
	&= \vartheta_i(t) - G(\mu, \gamma, \xi)(t-) + G(\mu, \gamma, \xi)(0) + \int_0^t \sum_{j=1}^d \vartheta_j(s)d\mu_j(s) + \sum_{0 \le s < t} \Delta G(\mu, \gamma, \xi)(s) \\
	&= \vartheta_i(t) + \Gamma^{G, c}(t),
	\end{align*}
	which is \eqref{eq : varphi disconti}. Here, the last identity follows from \eqref{eq : G1 disconti} and \eqref{eq : Gamma conti}. The equality $\sum_j \mu_j \equiv 1$, with \eqref{eq : varphi disconti} and \eqref{eq: value G}, now establishes the last claim
	\begin{align*}
	V^{\varphi}(t) &= \sum_{i=1}^d \varphi_i(t) \mu_i(t) = \sum_{i=1}^d \vartheta_i(t)\mu_i(t) + \Gamma^{G, c}(t) = G(\mu, \gamma, \xi)(t-) + \Gamma^{G, c}(t).
	\end{align*}
\end{proof}

\medskip

\begin{thm} [\textit{Multiplicative Functional Generation}]	\label{thm : multiplicative generation disconti}
	For a function $G:supp(\mu) \times supp(\gamma) \times supp(\xi) \rightarrow \mathbb{R}$ in $C^{2, 1, 1}$ such that the process $1/G(\mu, \gamma, \xi)$ is locally bounded and $G$ satisfies the balance condition \eqref{def : balanced mu}, we recall the vector process $\vartheta = (\vartheta_1, \cdots, \vartheta_d)'$ of \eqref{def : vartheta disconti} and define two vector processes $\eta = (\eta_1, \cdots, \eta_d)'$ and $\psi = (\psi_1, \cdots, \psi_d)'$ with components
	\begin{equation}		\label{def : eta disconti}
	\eta_i(t) := \vartheta_i(t) \cdot \exp\bigg(\int_0^t\frac{d\Gamma^{G, c}(s)}{G(\mu, \gamma, \xi)(s-)}\bigg), \qquad t \ge 0,
	\end{equation}
	\begin{equation}		\label{def : psi disconti}
	\psi_i(t) := \eta_i(t) - Q^{\eta, \mu}(t), \qquad t \ge 0,
	\end{equation}
	where $Q^{\eta, \mu}$ is defined in the manner of \eqref{def : defect of sf disconti}. Then $Q^{\eta, \mu}(\cdot) \equiv 0$ in \eqref{def : defect of sf disconti}, the trading strategy $\psi$ in \eqref{def : psi disconti} is represented for $i=1, \cdots, d$, as
	\begin{equation}	\label{eq : psi disconti}
	\psi_i(t) = \eta_i(t), \qquad t \ge 0,
	\end{equation}
	and generates the relative wealth process
	\begin{equation}	 \label{eq : MG wealth disconti}	
	V^{\psi}(t) =  G(\mu, \gamma, \xi)(t-) \cdot \exp\bigg(\int_0^t\frac{d\Gamma^{G, c}(s)}{G(\mu, \gamma, \xi)(s-)}\bigg), \qquad t \ge 0.
	\end{equation}
\end{thm}

\begin{proof}
	First, we can argue that $\psi$ is a trading strategy as in the proof of Theorem~\ref{thm : additive generation disconti}. We denote the continuous process
	\begin{equation*}
	K(t) := \exp\bigg(\int_0^t\frac{d\Gamma^{G, c}(s)}{G(\mu, \gamma, \xi)(s)}\bigg), \qquad t \ge 0,
	\end{equation*}
	and apply the product rule for jump processes along with \eqref{eq : G1 disconti}, \eqref{eq : Gamma conti}, \eqref{def : vartheta disconti}, and \eqref{def : eta disconti} to obtain
	\begin{align}
	d\Big(G(\mu, \gamma, \xi)(t)K(t)\Big) 
	&= K(t) \Big ( dG(\mu, \gamma, \xi)(t) + d\Gamma^{G, c}(t) \Big)	\label{eq: GK disconti}
	\\
	&= K(t) \sum_{i=1}^d D_{\mu_i}G(\mu, \gamma, \xi)(t-) d\mu_i(t) + K(t) \Delta G(\mu, \gamma, \xi)(t)									\nonumber
	\\
	&= \sum_{i=1}^d \eta_i(t) d\mu_i(t) + \Delta \Big(G(\mu, \gamma, \xi)(t) K(t)\Big)		\nonumber
	\\
	& = \sum_{i=1}^d \psi_i(t)d\mu_i(t) + \Delta \Big(G(\mu, \gamma, \xi)(t) K(t)\Big).			\nonumber	
	\end{align}
	The last equality used the definition \eqref{def : psi disconti} with the identity $\sum_j d\mu_j \equiv 0$. At $t=0$, we have
	\begin{align}
	V^{\psi}(0) 
	&= \sum_{i=1}^d \psi_i(0) \mu_i(0) 
	= \sum_{i=1}^d \eta_i(0)\mu_i(0)		\label{eq: GK at 0}
	\\
	&= \sum_{i=1}^d \vartheta_i(0)\mu_i(0)
	= G(\mu, \gamma, \xi)(0) = G(\mu, \gamma, \xi)(0)K(0),	\nonumber
	\end{align}
	where the second to last equality follows from \eqref{eq: value G}. Moreover, the trivial identity $\Delta Q^{\eta, \mu}(\cdot) \equiv 0$ with \eqref{eq: jump value G} yields for every $t \ge 0$
	\begin{align}	
	\Delta V^{\psi}(t) 
	&= \sum_{i=1}^d \Delta \psi_i(t)\mu_i(t)
	= \sum_{i=1}^d \Delta \eta_i(t)\mu_i(t)		\label{eq: GK jump}
	\\
	&= K(t) \sum_{i=1}^d \Delta \vartheta_i(t)\mu_i(t)
	= \Delta \Big(G(\mu, \gamma, \xi)(t)K(t)\Big).		\nonumber
	\end{align}
	Therefore, we arrive at \eqref{eq : MG wealth disconti}, from \eqref{eq: GK disconti}, \eqref{eq: GK at 0} and \eqref{eq: GK jump}.
	
	Finally, again from \eqref{eq: value G}, \eqref{eq: jump value G}, along with \eqref{eq: GK disconti}, we deduce
	\begin{align*}
	Q^{\eta, \mu}(t)
	&= K(t)\sum_{i=1}^d \vartheta_i(t)\mu_i(t) - K(0)\sum_{i=1}^d \vartheta_i(0)\mu_i(0) 
	\\
	& \qquad \qquad \qquad - \int_0^t \sum_{i=1}^d \eta_i(s)d\mu_i(s) + \sum_{0 \le s < t} K(s) \sum_{i=1}^d \Delta \vartheta_i(s)\mu_i(s)
	\\
	&= K(t)G(\mu, \gamma, \xi)(t-) - K(0)G(\mu, \gamma, \xi)(0) 
	\\
	& \qquad \qquad \qquad - \int_0^t \sum_{i=1}^d \psi_i(s)d\mu_i(s) - \sum_{0 \le s < t} \Delta G(\mu, \gamma, \xi)(s)K(s) = 0,
	\end{align*}
	and the identity \eqref{eq : psi disconti} follows.
\end{proof}

\medskip

We conclude with a note that it is hard to apply the method used in this section to rank based portfolios studied in Subsection~\ref{subsec: ranked}. When the components of $\Lambda$ is discontinuous, the dynamics of ranked functions $df_{[k]}(\mu, \Lambda)$ cannot be obtained in the form of \eqref{eq : ranked mu integral}, in terms of the original functions $f_j(\mu, \Lambda)$ and the collision local time terms; jumps of the components of $\Lambda$ may change the ranks between the components of $f(\mu, \Lambda)$ without accumulating the collision local times. Thus, for the examples of rank based portfolios which will be given in Subsection~\ref{subsec: CR by rank}, we assume that the components of $\Lambda$ have continuous paths.

\bigskip

\section{Portfolios depending on market-to-book ratios}	\label{sec: examples}

There is no restriction for investors on choosing the auxiliary process $\Lambda$ in the functional generation of trading strategies, thus any observable variables from the financial market, which they believe to be relevant in their portfolios, can be used as the components of $\Lambda$, as long as they are continuous for general generating functions as in Section~\ref{sec: FGP}, or RCLL for balanced generating functions as in Section~\ref{sec: disconti}. Some examples of such auxiliary variables are each company's expected profitability and expected investment, which are included in the five-factor asset pricing model \cite{Fama:French2006}, \cite{Fama:French2015}.

\smallskip

In this section, we adopt individual companies' book values as a primary example of the auxiliary process $\Lambda$, and construct portfolios which depend on the market-to-book ratios of the stocks. We write $b = (b_1, \cdots, b_d)'$ the vector representing companies' \textit{book values}, with each component $b_i$, $i = 1, \cdots, d$ an adapted, positive RCLL process of finite variation on compact time-intervals. The components of the \textit{relative book value} vector process $\beta = (\beta_1, \cdots, \beta_d)'$ are then defined as
\begin{equation}	\label{def : beta}
\beta_i(t) := \frac{b_i(t)}{\sum_{j=1}^d b_j(t)}, \qquad i = 1, \cdots, d, \qquad t \ge 0.
\end{equation}

\smallskip

We next introduce the vector process of \textit{market-to-book ratios} $\rho = (\rho_1, \cdots, \rho_d)'$ with components
\begin{equation}	\label{def : rho}
\rho_i := \frac{\mu_i}{\beta_i}, \qquad i = 1, \cdots, d, \qquad t \ge 0,
\end{equation}
which represent the market-to-book ratios of the various stocks in market. Here, we introduced \textit{relative} book values $\beta_i$ in \eqref{def : beta}, as the market weights $\mu_i$ of \eqref{def : market weights}, used to generate portfolios in earlier sections, represents the \textit{relative} capitalization, thus $\rho_i$ is a ratio between two \textit{relative} quantities. Moreover, since the sum of $\beta_i$'s is $1$ at all times, the vector $\beta$ itself is a portfolio, which appears in the following Example~\ref{ex: book value portfolio}. We note from the product rule for jump processes that the dynamics
\begin{equation*}
d\rho_i(t) = \frac{1}{\beta_i(t-)}d\mu_i(t) - \frac{\rho_i(t-)}{\beta_i(t-)}d\beta^c_i(t) + \Delta \rho_i(t), \qquad i = 1, \cdots, d.
\end{equation*}

\smallskip

Finally, we consider canonical decompositions
\begin{equation}		\label{eq : beta}
\beta_i := g_i - h_i, \qquad i = 1, \cdots, d,
\end{equation}	
where $g_i(\cdot)$ and $h_i(\cdot)$ are the smallest nondecreasing, adapted processes with $g_i(0) = \beta_i(0)$, for which \eqref{eq : beta} holds. We note that all component processes $\mu_i, \beta_i, \rho_i, g_i$ are positive, and $h_i$ are nonnegative with $h_i(0) = 0$ for $i = 1, \cdots, d$. We then let the generating function $G$ in $C^{2, 1, 1}$ depend on three processes $\rho, g, h$, by taking $\mu$, $\gamma \equiv g$, $\xi \equiv h$ as its arguments. 

\smallskip

All examples in this section, except Example~\ref{ex: logarithmic}, have balanced generating functions. Therefore, we assume that the book value processes $b_i$, thus also $\beta_i$, $\rho_i$, $g_i$, and $h_i$, have RCLL paths for $i = 1, \cdots, d$, except Example 5.5 and examples in Subsection~\ref{subsec: CR by rank}~(due to the difficulty addressed at the end of Section~\ref{sec: disconti}).

\medskip

\subsection{Multiplicatively generated portfolios}	\label{subsec: Examples of MGP}

We introduce in this subsection two important multiplicatively generated portfolios. 

\smallskip

\begin{example} [\textit{Book value portfolio}] \label{ex: book value portfolio}
	The vector process $\mu$ of market weights defined in \eqref{def : market weights} is itself a long-only portfolio, which invests the proportion $\mu_i(t)$ of current wealth in the $i$-th stock at all times. It is generated multiplicatively by the constant function, as seen in \cite{Fe}. In the same manner, we consider a long-only portfolio $\beta$ which invests the proportion $\beta_i(t-)$ of current wealth in the $i$-th stock at any time $t$. We call the latter \textit{book value portfolio}, and show in what follows that it can also be generated multiplicatively.
	
	\smallskip
	
	We consider the generating function
	\begin{equation}		\label{def : book value G}
	G(\mu, g, h)(t) := \frac{\prod_{i=1}^d \big(\frac{\mu_i(t)}{g_i(t) - h_i(t)}\big)^{g_i(t)-h_i(t)}}{\prod_{i=1}^d \big(\frac{\mu_i(0)}{g_i(0)}\big)^{g_i(0)}}
	=\frac{\prod_{i=1}^d \big(\rho_i(t)\big)^{\beta_i(t)}}{\prod_{i=1}^d \big(\rho_i(0)\big)^{\beta_i(0)}},
	\end{equation}
	and assume that the market-to-book ratios in \eqref{def : rho} satisfy
	\begin{equation}	\label{def : bounds of rho}
	m \le \rho_i(t) \le M, \qquad \forall~i = 1, \cdots, d, \qquad \forall~t \ge 0
	\end{equation}
	for some positive real constants $m < M$. Then the process $1/G(\rho(\cdot))$ is locally bounded. This assumption excludes the possibilities of overpricing by more than the factor $M$, and of underpricing by more than the factor $m$, for any stock in the market at any time. We then have 
	\begin{equation}	\label{eq : partial rho}
	D_{\mu_i}G(\mu, g, h) = \frac{\beta_i}{\mu_i}G(\mu, g, h), \qquad i = 1, \cdots, d,
	\end{equation}
	and deduce that $G$ is balanced as in \eqref{def : balanced mu}, thanks to the property $\sum_{i=1}^d \beta_i \equiv 1$. 
	
	\smallskip
	
	Moreover, we have from \eqref{def : eta disconti}, \eqref{eq : psi disconti} the quantities
	\begin{equation*}
	\eta_i(t) = \frac{G(\mu, g, h)(t-)}{\rho_i(t-)} \exp\bigg(\int_0^t\frac{d\Gamma^{G, c}(s)}{G(\mu, g, h)(s-)}\bigg), \qquad t \ge 0,
	\end{equation*}
	and the portfolio $\pi^{\psi}(\cdot)$ corresponding to the multiplicatively generated strategy $\psi = \eta$ is
	\begin{equation}		\label{def : book value portfolio}
	\pi^{\psi}_i(t) = \frac{\eta_i(t)\mu_i(t)}{\sum_{j=1}^d \eta_j(t) \mu_j(t)} = \beta_i(t-), \qquad i = 1, \cdots, d, \qquad t \ge 0.
	\end{equation}
	This portfolio indeed invests in the various assets, in proportion to their relative book values.
	
	We denote $V^{\beta}(\cdot) \equiv V^{\psi}(\cdot)$ the relative wealth process this portfolio generates. The log of this process can be computed from \eqref{eq : MG wealth disconti}, \eqref{eq : Gamma conti} in a manner devoid of stochastic integration:
	\begin{align}
	\log V^{\beta}(t) &= \log \bigg(\prod_{i=1}^d \big(\rho_i(t-)\big)^{\beta_i(t-)}\bigg) - \log \bigg(\prod_{i=1}^d \big(\rho_i(0)\big)^{\beta_i(0)}\bigg)		\label{G in BVP}
	\\
	& \qquad + \frac{1}{2}\int_0^t \sum_{i=1}^d \frac{d\langle \mu_i \rangle(s)}{\rho_i(s-)\mu_i(s)} -\frac{1}{2} \int_0^t \sum_{i, j = 1}^d \frac{d\langle \mu_i, \mu_j \rangle(s)}{\rho_i(s-)\rho_j(s-)}			\label{Quad in BVP}
	\\
	& \qquad -\int_0^t \sum_{i=1}^d \log\rho_i(s-) d\beta^c_i(s), \qquad t \ge 0.	\label{Beta term in BVP}
	\end{align}
	The empirical evolutions of the log relative process $\log V^{\beta}$ and the terms on the right-hand side will be given in Section~\ref{sec: empirical results}.
	
	\smallskip
	
	If an investor wants to use the book value portfolio $\beta$ as a benchmark, instead of the market portfolio $\mu$ of \eqref{def : market weights}, she can compute the relative value of the functionally generated trading strategies $\varphi$ and $\psi$ in Theorems~\ref{thm : additive generation disconti}, \ref{thm : multiplicative generation disconti} with respect to $\beta$, by dividing $V^{\varphi}$, $V^{\psi}$ of \eqref{eq : AG wealth disconti}, \eqref{eq : MG wealth disconti} into $V^{\beta}$, respectively.
\end{example}

\medskip

\begin{example} [\textit{Market-to-book ratio weighted portfolio}]	\label{ex: market-to-book weighted portfolio}
	It is well-known that the so-called ``diversity-weighted'' portfolio $\pi^{(p)} = (\pi^{(p)}_1, \cdots, \pi^{(p)}_d)'$ with components
	\begin{equation}	\label{def : diversity weighted portfolio}
	\pi^{(p)}_i(t) := \frac{\big(\mu_i(t)\big)^p}{\sum_{j=1}^d \big(\mu_j(t)\big)^p}, \qquad i = 1, \cdots, d, \qquad t \ge 0,
	\end{equation}
	and $0 < p < 1$, is strong arbitrage relative to the market under mild conditions. See, for example, Section~6.2 of \cite{Fe}, or Section~7 of \cite{FK_survey}. This outperformance is due to the underweighting of the larger stocks and the overweighting of the smaller stocks relative to the market, by the portfolio of \eqref{def : diversity weighted portfolio}. Similar results can be found in \cite{Vervuurt:Karatzas:2015} for negative exponents $p < 0$.
	
	\smallskip
	
	In a similar vein, we consider the portfolio $\Pi^{(p)} = (\Pi^{(p)}_1, \cdots, \Pi^{(p)}_d)'$ with components
	\begin{equation}	\label{def : MTBR weighted portfolio}
	\Pi^{(p)}_i(t) := \frac{\big(\rho_i(t-)\big)^p}{\sum_{j=1}^d \big(\rho_j(t-)\big)^p}, \qquad i = 1, \cdots, d, \qquad t \ge 0.
	\end{equation}
	It is straightforward from \eqref{eq : psi disconti} and \eqref{def : portfolio} that $\Pi^{(p)}$ is multiplicatively generated from the balanced function
	\begin{equation}		\label{def: diversity-weighted}
	G(\mu, g, h)(t) := \Bigg(\frac{\sum_{i=1}^d \Big(\frac{\mu_i(t)}{g_i(t) - h_i(t)}\Big)^p} {\sum_{i=1}^d \Big(\frac{\mu_i(0)}{g_i(0)}\Big)^p} \Bigg)^{1/p}
	= \bigg(\frac{\sum_{i=1}^d \rho_i^p(t)}{\sum_{i=1}^d \rho_i^p(0)}\bigg)^{1/p},
	\end{equation}
	which is concave for $p < 1$. For these values of $p$, the portfolio tends to invest higher proportions of capital in the stocks with smaller market-to-book ratios. This tilt toward the stocks with small market-to-book ratios becomes more severe as $p$ gets smaller. The empirical results of $\Pi^{(p)}$ with different values of $p$ will be given in Section~\ref{sec: empirical results}.
\end{example}

\medskip

\subsection{Additively generated portfolios}		\label{subsec: Examples of AGP}

We introduce now additive analogues of the portfolios in Example~\ref{ex: book value portfolio} and \ref{ex: market-to-book weighted portfolio}. Our primary purpose here, is to make the Gamma process nondecreasing, in order to find the smallest time-interval $[0, T_*]$ as in \eqref{con : AGSRA mu}, over which the generated portfolio outperforms the market with probability 1.

\smallskip

\begin{example} [\textit{Modified book value portfolio}] \label{ex: modified book-value}
	The integral in \eqref{Beta term in BVP} involves the non-monotone integrator $d\beta_i^c$, which makes finding strong relative arbitrage hard. In this endeavor, we consider the slight modification without normalization
	\begin{equation}	\label{def : book value G2}	
	G(\mu, g, h) := \prod_{i=1}^d \bigg(\Big(\frac{\mu_i}{g_i-h_i}\Big)^{g_i-h_i} \cdot \exp\Big[(1-\log M)g_i + (\log m -1)h_i\Big]\bigg)	
	\end{equation}
	of the function in \eqref{def : book value G}, with the bounds for $\rho$ in \eqref{def : bounds of rho} and $g_i > h_i \ge 0$. For this function, we have again the properties \eqref{eq : partial rho} and \eqref{def : balanced mu}, as well as
	\begin{align*}
	D_{g_i}G(\mu, g, h) &= \Big(\log \frac{\rho_i}{M}\Big)G(\mu, g, h)
	\\
	D_{h_i}G(\mu, g, h) &= \Big(\log \frac{m}{\rho_i}\Big)G(\mu, g, h), \qquad i = 1, \cdots, d.
	\end{align*}
	In particular, this function satisfies the requirements \eqref{Eq : two inequalities mu}. The expression of \eqref{eq : Gamma conti} becomes now
	\begin{align}	\label{eq : gamma G2}
	\Gamma^{G, c}(t) = &- \frac{1}{2} \sum_{i, j=1}^d \int_0^t D^2_{\mu_i, \mu_j} G(\mu, g, h)(s-) d\langle \mu_i, \mu_j \rangle (s)
	\\
	& + \sum_{i=1}^d \int_0^t G(\mu, g, h)(s-) \bigg[ \log \Big(\frac{M}{\rho_i(s-)}\Big) dg^c_i(s) + \log \Big(\frac{\rho_i(s-)}{m}\Big) dh^c_i(s) \bigg],	\nonumber
	\end{align}
	and is clearly nondecreasing since the function $G(\cdot, g, h)$ is concave.
	
	\smallskip
	
	The processes $\vartheta_i(\cdot)$ of \eqref{def : vartheta disconti} are
	\begin{equation}	\label{eq : vartheta of modified book value portfolio}
	\vartheta_i(t) = \frac{G(\mu, g, h)(t-)}{\rho_i(t-)}, \qquad i = 1, \cdots, d, \qquad t \ge 0,
	\end{equation}
	and thus the additively generated trading strategy $\varphi$ is obtained from \eqref{eq : varphi disconti} as
	\begin{equation}	\label{eq : phi of modified book value portfolio}
	\varphi_i(t) = \frac{G(\mu, g, h)(t-)}{\rho_i(t-)} + \Gamma^{G, c}(t), \qquad i = 1, \cdots, d, \qquad t \ge 0.
	\end{equation}
	This trading strategy $\varphi(t)$ holds positive positions in all assets at all times. It invests the baseline amount $\Gamma^{G, c}(t)$ across all stocks, and the quantity $G(t-)$ discounted by the reciprocal $1/\rho_i(t-)$ of the market-to-book ratio for the $i$-th stock. Thus, $\varphi$ invests less capital in the stocks of companies with higher market-to-book ratios, and generates wealth given by \eqref{eq : AG wealth disconti}.
	
	\smallskip
	
	Proposition~\ref{prop : AGSRA mu} shows that the strategy $\varphi$ is strong relative arbitrage with respect to the market over every time-interval $[0, t]$ with $t \ge T_*$, for any $T_*$ satisfying
	\begin{equation*}
	\Gamma^{G, c}(T_*) \ge \prod_{i=1}^d \big(\rho_i(0)e^{(1-\log M)}\big)^{\beta_i(0)}, \qquad \mathbb{P}-a.e.
	\end{equation*}
	The portfolio $\pi^{\varphi}$ corresponding to the strategy $\varphi$, is computed from \eqref{def : portfolio} as
	\begin{equation}		\label{eq: modified book value portfolio}	
	\pi^{\varphi}_i(t) = \frac{\beta_i(t-)G(\mu, g, h)(t-) + \mu_i(t)\Gamma^{G, c}(t)}{G(\mu, g, h)(t-) + \Gamma^{G, c}(t)}, \qquad i = 1, \cdots, d, \qquad t \ge 0.
	\end{equation}
	At any given time $t \ge 0$, this portfolio interpolates linearly between the market portfolio $\mu(t)$ of \eqref{def : market weights} (with weight $\Gamma^{G, c}(t)/(G(\mu, g, h)(t-)+\Gamma^{G, c}(t))$), and the book-value portfolio $\beta(t-)$ of \eqref{def : beta} (with weight $G(\mu, g, h)(t-)/(G(\mu, g, h)(t-)+\Gamma^{G, c}(t))$).
	
	\smallskip
	
	We note that the portfolio weights multiplicatively generated by the function \eqref{def : book value G2} coincides with the book-value portfolio \eqref{def : book value portfolio}, from \eqref{eq : psi disconti}, \eqref{def : eta disconti}, \eqref{eq : vartheta of modified book value portfolio}, and \eqref{def : portfolio}.
\end{example}

\medskip

We assume in the following examples that the relative book values $\beta_i$ admit a universal lower bound $\delta > 0$, i.e.,
\begin{equation}		\label{def : delta}
\delta \le \beta_i(t), \qquad \forall~i = 1, \cdots, d, \qquad \forall~t \ge 0.
\end{equation}

\smallskip

\begin{example}	[\textit{Modified market-to-book ratio weighted portfolio}]
	In order to achieve a nondecreasing Gamma process, we consider the generating function for any $p \neq 0$
	\begin{equation}	\label{def: modified diversity-weighted}
	G(\mu, g, h)(t) := \bigg(\sum_{i=1}^d \Big(\frac{\mu_i(t)g_i(t)}{g_i(t)-h_i(t)} e^{-\frac{h_i(t)}{\delta}}\Big)^p\bigg)^{1/p}
	= \bigg(\sum_{i=1}^d \Big(\rho_i(t)g_i(t) e^{-\frac{h_i(t)}{\delta}}\Big)^p\bigg)^{1/p},
	\end{equation}
	a modification of \eqref{def: diversity-weighted}. It is easily checked that the function $G$ of \eqref{def: modified diversity-weighted} satisfies the condition \eqref{Eq : two inequalities mu}, and that the Gamma process of \eqref{eq : Gamma conti} is nondecreasing, and computed as	
	\begin{align}
	\Gamma^{G, c}(t) = &- \frac{1}{2} \int_0^t \sum_{i, j = 1}^d D^2_{\mu_i, \mu_j} G(\mu, g, h)(s-) d\langle \mu_i, \mu_j \rangle (s)	\label{eq : Gamma modified diversity-weighted}
	\\
	& + \int_0^t \sum_{i=1}^d G(\mu, g, h)(s-) \bigg( \frac{h_i(s-)}{\beta_i(s-)g_i(s-)} \bigg) \frac{\big(\rho_i(s-)g_i(s-)e^{-\frac{h_i(s-)}{\delta}}\big)^p}{\sum_{j=1}^d \big(\rho_j(s-)g_j(s-)e^{-\frac{h_j(s-)}{\delta}}\big)^p} dg^c_i(s)		\nonumber
	\\
	& + \int_0^t \sum_{i=1}^d G(\mu, g, h)(s-) \bigg( \frac{1}{\delta} - \frac{1}{\beta_i(s-)} \bigg) \frac{\big(\rho_i(s-)g_i(s-)e^{-\frac{h_i(s-)}{\delta}}\big)^p}{\sum_{j=1}^d \big(\rho_j(s-)g_j(s-)e^{-\frac{h_j(s-)}{\delta}}\big)^p} dh^c_i(s), \qquad t \ge 0.			\nonumber
	\end{align}
	Thus, the condition \eqref{con : AGSRA mu} of Proposition~\ref{prop : AGSRA mu}, with $\Gamma^G(T_*)$ replaced by $\Gamma^{G, c}(T_*)$, leads to strong relative arbitrage.
	From \eqref{eq : varphi disconti} and \eqref{def : portfolio}, the portfolio $\pi^{\varphi}$ corresponding to the additively generated strategy $\varphi$, is computed as
	\begin{equation*}
	\pi^{\varphi}_i(t) = \frac{\widehat{\pi}_i(t-) G(\mu, g, h)(t-) + \mu_i(t) \Gamma^{G, c}(t)}{G(\mu, g, h)(t-) + \Gamma^{G, c}(t)}.
	\end{equation*}
	This portfolio is again a linear interpolation between the market portfolio $\mu$, and the portfolio $\widehat{\pi}$ with components
	\begin{equation*}
	\widehat{\pi}_i(t) = \frac{\big(\rho_i(t) g_i(t) e^{-\frac{h_i(t)}{\delta}}\big)^p}{\sum_{j=1}^d \big(\rho_j(t) g_j(t) e^{-\frac{h_j(t)}{\delta}}\big)^p} \qquad i = 1, \cdots, d, \qquad t \ge 0.
	\end{equation*}
	
	\smallskip
	
	When $p = 1$, the function $G$ of \eqref{def: modified diversity-weighted} is linear in $\mu$, and the first integral on the right-hand side of \eqref{eq : Gamma modified diversity-weighted} vanishes, thus the growth of the Gamma process depends only on the growth of $g_i$ and $h_i$.
\end{example}

\medskip

We continue with one more example of additively-generated strong arbitrage relative to the market portfolio $\mu$ of \eqref{def : market weights}.

\smallskip

\begin{example} [\textit{Logarithmic generating function}]	\label{ex: logarithmic}
	In this example, the generating function does not satisfy the balanced condition so we assume relative book values $\beta_i(\cdot)$ to be continuous. We assume further the lower bound $\delta$ for $\beta_i(\cdot)$ in \eqref{def : delta}, as well as the bounds $m$, $M$ for market-to-book ratios $\rho_i(\cdot)$ in \eqref{def : bounds of rho}.
	We then consider the generating function
	\begin{equation*}
	G(\mu, g, h) := \sum_{i=1}^d \Big(\log\big(1+\frac{\mu_i}{g_i-h_i}\big)\Big) \exp \bigg(-\frac{h_i}{\delta\kappa} \bigg)
	= \sum_{i=1}^d \Big(\log(1+\rho_i)\Big) \exp \bigg(-\frac{h_i}{\delta\kappa} \bigg),
	\end{equation*}
	which is concave in the argument $\mu$, for the positive constant
	\begin{equation*}
	\kappa := \frac{1+m}{m}\log(1+m).
	\end{equation*}
	Since the function 
	\begin{equation*}
	f(x) := \frac{1+x}{x}\log(1+x)
	\end{equation*}
	is nondecreasing, we have $\kappa \le f(\rho_i(t))$ for all $i=1, \cdots, d$, and $t \ge 0$. Computations show that $G$ satisfies the inequalities \eqref{Eq : two inequalities mu} and admits the nondecreasing Gamma process
	\begin{align*}
	\Gamma^G(t) & = \frac{1}{2} \int_0^t \sum_{i=1}^d \frac{1}{\big(\beta_i(s)+\mu_i(s)\big)^2}\exp\Big(-\frac{h_i(s)}{\delta\kappa}\Big) d\langle \mu_i \rangle (s)		
	\\
	& \quad + \int_0^t \sum_{i=1}^d \frac{\rho_i(s)}{\beta_i(s)+\mu_i(s)}\exp\Big(-\frac{h_i(s)}{\delta\kappa}\Big) dg_i(s)
	\\
	& \quad + \int_0^t \sum_{i=1}^d \bigg( \frac{\log(1+\rho_i(s))}{\delta\kappa}-\frac{\rho_i(s)}{\beta_i(s)+\mu_i(s)} \bigg) \exp\Big(-\frac{h_i(s)}{\delta\kappa}\Big) dh_i(s).
	\end{align*}
	from \eqref{eq : Gamma2 mu}. In Proposition~\ref{prop : AGSRA mu}, the trading strategy $\varphi$, additively generated from $G$, is a strong relative arbitrage with respect to the market over any time-interval $[0, t]$ with $t \ge T_*$, satisfying
	\begin{equation*}
	\Gamma^{G}(T_*) > G(\mu, g, h)(0) = \sum_{i=1}^d \log\big(1+\rho_i(0)\big), \qquad \mathbb{P}-a.e.
	\end{equation*}
\end{example}

\medskip

\subsection{Constant rebalanced portfolio by ranks of market-to-book ratios}	\label{subsec: CR by rank}

We present in this subsection a class of portfolios depending on the ranks of market-to-book ratios. In the setting of Example~\ref{ex : ranked market-to-book ratio}, we write $\bm{\rho} \equiv \pazocal{R}\big(f(\mu, \Lambda)\big) \equiv \bm{\nu}$ the ranked market-to-book ratio vector with components $\bm{\rho} = (\rho_{[1]}, \rho_{[2]}, \cdots, \rho_{[d]})'$ in descending ranks. Then the vector $\vartheta$ of \eqref{def : ranked vartheta mu} and the Gamma process of \eqref{eq : ranked Gamma mu} take simplified forms, as
\begin{equation}	\label{def : ranked vartheta}
\vartheta_i(t) := \sum_{k=1}^d \frac{1}{\beta_i(t)} D_{k}G\big(\bm{\rho}(t)\big) I_{\{r_t(k)=i\}}, \qquad t \ge 0,
\end{equation}
and
\begin{align}
\Gamma^G(t) & = \int_0^t \sum_{k, i=1}^d D_{k}G\big(\bm{\rho}(s)\big) I_{\{r_s(k)=i\}} \frac{\rho_i(s)}{\beta_i(s)} d\beta_i(s) 		\label{eq : ranked Gamma}
\\
&~~~-\frac{1}{2} \int_0^t \sum_{k, \ell, i, j=1}^d D_{k, \ell}^2 G\big(\bm{\rho}(s)\big) \frac{I_{\{r_s(k)=i\}} I_{\{r_s(\ell)=j\}}}{\beta_i(s)\beta_j(s)} d\langle \mu_i, \mu_j \rangle(s)	\nonumber
\\
&~~~-\frac{1}{2}\int_0^t \sum_{k=1}^d D_{k}G\big(\bm{\rho}(s)\big)dL^{\rho_{[k]}-\rho_{[k+1]}}(s)
+\frac{1}{2}\int_0^t \sum_{k=1}^d D_{k}G\big(\bm{\rho}(s)\big)dL^{\rho_{[k-1]}-\rho_{[k]}}(s).	\nonumber
\end{align}

\smallskip

Let us fix a vector $c = (c_1, \cdots, c_d)'$ of constants in $\mathbb{R}^d$, satisfying $\sum_{k=1}^d c_k = 1$, and consider the function 
\begin{equation}	\label{eq: CR function}
G\big(\bm{\rho}(t)\big) = \prod_{k=1}^d \big( \rho_{[k]}(t) \big)^{c_k}.
\end{equation}
Basic computations show that the vector process $\vartheta$ in \eqref{def : ranked vartheta}, the multiplicatively generated trading strategy $\psi$ of \eqref{eq : psi s}, and its corresponding portfolio of \eqref{def : portfolio}, are obtained respectively as
\begin{align}
\vartheta_i(t) &= \sum_{k=1}^d \frac{c_k I_{\{r_t(k)=i\}}}{\beta_i(t)\rho_{[k]}(t)}G\big(\bm{\rho}(t)\big), 	\nonumber \\
\psi_i(t) &= \exp\bigg(\int_0^t\frac{d\Gamma^{G}(s)}{G\big(\bm{\rho}(s)\big)}\bigg) \sum_{k=1}^d \frac{c_k I_{\{r_t(k)=i\}}}{\beta_i(t)\rho_{[k]}(t)}G\big(\bm{\rho}(t)\big),	\nonumber \\
\pi^{\psi}_i(t) &= \sum_{k=1}^d c_k I_{\{r_t(k)=i\}},		\label{eq: CR portfolio}
\end{align}
for $i=1, \cdots d$. Thus, this portfolio $\pi^{\psi}$ rebalances at all times to maintain a constant proportion $c_k$ of current wealth invested in the stock having $k$-th ranked market-to-book ratio, for $k \le n$. By choosing the exponent vector $c$ in \eqref{eq: CR function}, we can construct several portfolios as the following. 

\medskip

\begin{example} [Equally-weighted portfolios composed of a fixed number of value stocks/growth stocks]	\label{ex: EW}
	We assume in this example that the number of stocks $d$ in the stock market is bigger than a fixed number $\ell \in \mathbb{N}$ and consider the vector $c$ with components
	\begin{equation*}
	c_k = 
	\begin{cases}
	& \frac{1}{\ell}, \qquad \text{for } 1 \le k \le \ell, \\
	& \, 0 ,  \qquad \text{for } \ell < k \le d.
	\end{cases}
	\end{equation*}
	The portfolio $\tilde{\pi}^{\psi}$ of \eqref{eq: CR portfolio} is then
	\begin{equation}	\label{def : EW top100}
	\tilde{\pi}^{\psi}_i(t) = \sum_{k=1}^{\ell} \frac{I_{\{r_t(k)=i\}}}{\ell}, \qquad i = 1, \cdots, d.
	\end{equation}
	This portfolio invests equal weights at all times to the top $\ell$ `value stocks', i.e., $\ell$ stocks with the highest market-to-book ratios.
	
	On the other hand, if we set the last $\ell$ components of $c$ to be $1/\ell$:
	\begin{equation*}
	c_k = 
	\begin{cases}
	& 0, \qquad \text{for } 1 \le k \le d-\ell, \\
	& \frac{1}{\ell}, \qquad \text{for } d-\ell < k \le d,
	\end{cases}
	\end{equation*}
	the corresponding portfolio 
	\begin{equation}	\label{def : EW low100}
	\underaccent{\tilde}{\pi}^{\psi}_i(t) = \sum_{k=d-\ell+1}^{d} \frac{I_{\{r_t(k)=i\}}}{\ell}, \qquad i = 1, \cdots, d,
	\end{equation}
	distributes equal weights to the top $\ell$ `growth stocks', i.e., the ones with the lowest market-to-book ratios, at all times.
\end{example}

\medskip

\begin{example} [The stock with the biggest/smallest market-to-book ratio; leakage]
	We start with the extreme case, where the vector $c$ in Example~\ref{ex: EW} is given as the unit vector $e_1 = (1, 0, \cdots, 0) \in \mathbb{R}^d$, and consider the generating function 
	\begin{equation*}
	G\big(\bm{\rho}(t)\big) = \rho_{[1]}(t).
	\end{equation*}
	Basic computations show that the multiplicatively generated trading strategy $\psi$ of \eqref{eq : psi s}, and its corresponding portfolio of \eqref{def : portfolio}, are obtained respectively as
	\begin{equation*}
	\psi_i(t) = \frac{I_{\{r_t(1)=i\}}}{\beta_i(t)} \cdot \exp \bigg( \int_0^t \sum_{i=1}^d \frac{I_{\{r_s(1)=i\}}}{\beta_i(s)}d\beta_i(s)-\frac{1}{2}\int_0^t \frac{dL^{\rho_{[1]}-\rho_{[2]}}(s)}{\rho_{[1]}(s)} \bigg),
	\end{equation*}
	\begin{equation*}
	\pi^{\psi}_i(t) = I_{\{r_t(1)=i\}},
	\end{equation*}
	for $i=1, \cdots d$. Thus, this portfolio invests only in the stock with the biggest market-to-book ratio at all times. Its log relative wealth process from \eqref{eq : MG wealth s} is then
	\begin{equation*}
	\log V^{\psi}(t) = \log \big( \rho_{[1]}(t) \big) + \int_0^t \sum_{i=1}^d \frac{I_{\{r_s(1)=i\}}}{\beta_i(s)}d\beta_i(s)-\frac{1}{2}\int_0^t \frac{dL^{\rho_{[1]}-\rho_{[2]}}(s)}{\rho_{[1]}(s)}.
	\end{equation*}
	The last nonincreasing term involving the local time captures the ``leakage effect'', caused by crossover from the stock with the biggest market-to-book ratio to the rest of the market. The second-last term, however, is not monotone and its direction coincides with that of the book value of the stock with the largest market-to-book ratio. The existence of such non-monotone term in the drift part of the log relative wealth, is a crucial difference to the case of the stock with the biggest size~(capitalization) in Example~4.3.1 of \cite{Fe}.
	
	\medskip
	
	In a similar manner, the unit vector $c = e_d = (0, \cdots, 0, 1) \in \mathbb{R}^d$ with
	\begin{equation*}
	G\big(\bm{\rho}(t)\big) = \rho_{[d]}(t)
	\end{equation*}
	generates multiplicatively the trading strategy $\psi$ with components
	\begin{equation*}
	\psi_i(t) = \frac{I_{\{r_t(d)=i\}}}{\beta_i(t)} \cdot \exp \bigg( \int_0^t \sum_{i=1}^d \frac{I_{\{r_s(d)=i\}}}{\beta_i(s)}d\beta_i(s)+\frac{1}{2}\int_0^t \frac{dL^{\rho_{[d-1]}-\rho_{[d]}}(s)}{\rho_{[d]}(s)} \bigg),
	\end{equation*}
	as well as the corresponding portfolio $\pi^{\psi}$ with components
	\begin{equation*}
	\pi^{\psi}_i(t) = I_{\{r_t(d)=i\}},
	\end{equation*}
	for $i=1, \cdots d$. This portfolio invests only in the stock with the smallest market-to-book ratio at all times, and its log relative wealth process is given as
	\begin{equation*}
	\log V^{\psi}(t) = \log \big( \rho_{[d]}(t) \big) + \int_0^t \sum_{i=1}^d \frac{I_{\{r_s(d)=i\}}}{\beta_i(s)}d\beta_i(s)+\frac{1}{2}\int_0^t \frac{dL^{\rho_{[d-1]}-\rho_{[d]}}(s)}{\rho_{[d]}(s)}.
	\end{equation*}
	The drift part again contains the non-monotone integral with the integrators $\beta_i$'s, and the nondecreasing integral which represents crossover from the stock with the smallest market-to-book ratio to the rest of the market.
\end{example}

\bigskip

\section{Market-to-book ratio component of portfolio returns}	\label{sec: PTB component}

In this section, we discuss the market-to-book ratio component of portfolio returns, which was introduced in Subsection 7.4 of \cite{Fe}. First, we present the method of measuring the component of relative portfolio returns, due to changes in the capital distribution.

\medskip

We define the $k$-th \textit{rank process} of the market weight vector $\mu$ by
\begin{equation}		\label{def : ranked mu}
\mu_{(k)}(t) := \max_{1 \le i_1 < \cdots < i_k \le d} \min \big(\mu_{i_1}(t), \cdots, \mu_{i_k}(t)\big), \qquad t \ge 0
\end{equation}
for $k = 1, \cdots, d$, so that we obtain the ranks of $\mu$ in descending order
\begin{equation*}
\mu_{(1)}(t) \ge \mu_{(2)}(t) \ge \cdots \ge \mu_{(d)}(t), \qquad t \ge 0,
\end{equation*}
with the same lexicographic rule for breaking ties as before. We denote by $\bm{\mu} = (\mu_{(1)}, \cdots, \mu_{(d)})'$ the vector of the ranked market weights. We also consider the random permutation process $p_t$ of $\{1, \cdots, d\}$ such that for $k = 1, \cdots, d$,
\begin{equation}		\label{def : permutation of mu}
\mu_{p_t(k)}(t) = \mu_{(k)}(t),
\end{equation}
\begin{equation*}
p_t(k) < p_t(k+1) \qquad \text{if} \qquad \mu_{(k)}(t) = \mu_{(k+1)}(t)
\end{equation*}
hold for every $t \ge 0$. Thus, $p_t(k)$ indicates the index name of the $k$-th ranked market weight $\mu_{(k)}$ at time $t$.

\medskip

For any given portfolio $\pi=(\pi_1, \cdots, \pi_d)'$, we define the vector process $w = (w_1, \cdots, w_d)'$ with components
\begin{equation}		\label{def : weight ratio by size}
w_k(t) := \frac{\pi_{p_t(k)}(t)}{\mu_{(k)}(t)}, \qquad t \ge 0, \qquad k = 1, \cdots, d,
\end{equation}
and call it the vector of \textit{weight ratios} for $\pi$. The quantity $w_k(t)$ represents the (fractional) number of shares of the $k$-th ranked stock held by the portfolio $\pi$ at time $t$. For a fixed time $t_0 \ge 0$, we consider the weight ratios $w_1(t_0), \cdots, w_d(t_0)$ of $\pi$ at $t_0$ and the function
\begin{equation*}
G\big(\bm{\mu}(t)\big) = w_1(t_0)\mu_{(1)}(t) + \cdots + w_d(t_0)\mu_{(d)}(t).
\end{equation*}
From Theorem~4.2.1 of \cite{Fe}, this function $G$ generates (multiplicatively) a new portfolio $\nu$ with components
\begin{equation*}
\nu_{p_t(k)}(t) = \frac{w_k(t_0)\mu_{(k)}(t)}{w_1(t_0)\mu_{(1)}(t)+\cdots+w_d(t_0)\mu_{(d)}(t)}, \qquad t \ge t_0
\end{equation*}
for $k=1, \cdots, d$, whose log relative wealth is given as
\begin{equation}		\label{eq : wealth of nu}
d\log V^{\nu}(t) = d\log G\big(\bm{\mu}(t)\big) + \frac{1}{2G\big(\bm{\mu}(t)\big)}\sum_{k=1}^{d-1}\big(w_{k+1}(t_0)-w_k(t_0)\big)dL^{\mu_{(k)}-\mu_{(k+1)}}(t), \qquad t \ge t_0.
\end{equation}
The weight ratios for $\nu$ are 
\begin{equation*}
\frac{\nu_{p_t(k)}(t)}{\mu_{(k)}(t)} = \frac{w_k(t_0)}{w_1(t_0)\mu_{(1)}(t)+\cdots+w_d(t_0)\mu_{(d)}(t)}, \qquad t \ge t_0,
\end{equation*}
proportional to the constants $w_k(t_0)$, for $k=1, \cdots, d$. In other words, the new portfolio $\nu$ keeps the number of shares of the $k$-th ranked stocks proportional to the constant $w_k(t_0)$ for $k = 1, \cdots, d$. Note that $G\big(\bm{\mu}(t_0)\big) = \sum_{k=1}^d \pi_{p_t(k)}(t_0) = 1$ and thus the two portfolios $\nu$ and $\pi$ coincide at time $t_0$.

\smallskip

Over a short period of time from $t_0$ to $t_1>t_0$, we estimate the first term on the right-hand side of \eqref{eq : wealth of nu} at $t_0$ as
\begin{align}
d\log G\big(\bm{\mu}(t_0)\big) 
&\approx \log G\big(\bm{\mu}(t_1)\big) - \log G\big(\bm{\mu}(t_0)\big)		\label{def : DC}
\\
&= \log \big(w_1(t_0)\mu_{(1)}(t_1)+\cdots+w_d(t_0)\mu_{(d)}(t_1)\big) =: DC_{\pi}(t_0:t_1).		\nonumber
\end{align}
We call this last term the \textit{distributional component} of the relative return over the period from $t_0$ and $t_1$. Since $\nu(t_0) = \pi(t_0)$, this component measures the effect on the relative return of both portfolios $\nu$ and $\pi$ caused by change in the capital distribution at time $t_0$. From a practical point of view, investors rebalance portfolios in discrete time so that the above approximation is justified when we pick $t_1$ smaller than the next time of rebalancing after $t_0$.

\medskip

We now adopt the same methodology to measure the component of relative portfolio returns due to changes in the market-to-book ratios. Let us recall the random permutation process $r_t$ in Example~\ref{ex : ranked market-to-book ratio} and denote $\widetilde{\bm{\mu}} = (\mu_{r_t(1)}, \cdots, \mu_{r_t(d)})'$ the vector of market weights arranged in descending order of market-to-book ratios.

\smallskip

For any given portfolio $\pi=(\pi_1, \cdots, \pi_d)'$, we consider the vector of weight ratios $v = (v_1, \cdots, v_d)'$ with components
\begin{equation}	\label{def : weight ratio by pbr}
v_k(t) := \frac{\pi_{r_t(k)}(t)}{\mu_{r_t(k)}(t)}, \qquad t \ge 0, \qquad k = 1, \cdots, d.
\end{equation}
The quantity $v_k(t)$ represents the (fractional) number of shares of the stock with the $k$-th highest market-to-book ratio held in $\pi$ at time $t$. We then define the \textit{market-to-book ratio component} of the relative return of $\pi$ over the period from $t_0$ to $t_1 > t_0$ to be
\begin{equation*}
\log \big(v_1(t_0)\mu_{r_{t_1}(1)}(t_1)+\cdots+v_d(t_0)\mu_{r_{t_1}(d)}(t_1)\big) =: MBRC_{\pi}(t_0:t_1).
\end{equation*}

\medskip

Therefore, if a given portfolio $\pi$ is rebalanced at discrete time points $t_0, t_1, t_2, \cdots$, we can compute two different vectors $w$ and $v$ of weight ratios in \eqref{def : weight ratio by size} and \eqref{def : weight ratio by pbr} which are arranged according to size, and according to market-to-book ratio, respectively, at each time $t_0, t_1, \cdots$. Then the quantities $DC_{\pi}(t_0:t_1), DC_{\pi}(t_1:t_2), \cdots$ in \eqref{def : DC} measure the effect on the relative return of $\pi$ due to change in size over time, whereas $MBRC_{\pi}(t_0:t_1), MBRC_{\pi}(t_1:t_2), \cdots$ represent the effect due to change in market-to-book ratio over time.

\bigskip

\section{Empirical results}	\label{sec: empirical results}

We present in this section some empirical results for the portfolios depending on the relative book values, or the market-to-book ratios, introduced in Section~\ref{sec: examples}. For some of the examples of portfolios $\pi$ studied in earlier sections, we provide their relative values $V^{\pi}$ with respect to the market. By examining these values, we show how the book values (or market-to-book ratios) affect the growth of portfolios.

\medskip

\subsection{Data description}
We used daily closing prices and year-end annual book values for 20 years (5032 consecutive trading days, between 2001 January 1st and 2020 December 31st). These data were obtained from the Compustat North America database, accessed via the Wharton research data services: \\ \href{https://wrds-www.wharton.upenn.edu/pages/get-data/compustat-capital-iq-standard-poors/}{https://wrds-www.wharton.upenn.edu/pages/get-data/compustat-capital-iq-standard-poors/}. 

\smallskip

As the number of companies in our dataset varies by each day, we selected $d = 500$ largest stocks in capitalization on the first trading day~(2001 January 1st) to construct the portfolios. Because such selection is subject to survivorship bias, we do not emphasize how much the portfolios outperform with respect to the market; we shall try to identify the value factor in portfolio returns by analyzing and comparing between the portfolios. We also note that no empirical results of the portfolios, which appeared in Subsection~\ref{subsec: Examples of AGP}, will be given, since they are designed to outperform the market. Moreover, their growth~(of the Gamma process) depends on the growth of $g$ and $h$ in the decomposition \eqref{eq : beta} of the relative book values $\beta$, but they remain constant most of the time in our dataset, since the book values are only updated annually. The empirical analysis of such portfolios will be meaningful if we have a dataset in which book values are updated more frequently.

\smallskip

For $N=5032$ trading days, we discretized the time horizon as $0=t_0 < t_1 < \cdots < t_{N-1} = T$ and computed the following values for each stock $i = 1, \cdots, 500$ for each day $\ell = 0, 1, \cdots, N-1$;
\begin{enumerate} [wide, labelwidth=!, labelindent=0pt]
	\item $S_i(t_\ell)$ : the capitalization (daily closing price multiplied by total number of outstanding shares) of $i$-th stock at the end of day $t_\ell$.
	
	\item $\Sigma(t_\ell) := \sum_{i=1}^d S_i(t_\ell)$ : the total capitalization of $d=500$ stocks at the end of day $t_\ell$. This quantity represents dollar value of the market portfolio at the end of day $t_\ell$.
	
	\item $\mu_i(t_\ell) := \frac{S_i(t_\ell)}{\Sigma(t_\ell)} $ : the $i$-th market weight at the end of day $t_\ell$, as defined in \eqref{def : market weights}.
	
	\item $\beta_i(t_\ell)$ : the relative book value of $i$-th stock, defined in \eqref{def : beta}, on day $t_\ell$. These values are updated annually, so they remain constant within every calendar year.
	
	\item $\rho_i(t_\ell)$ : the market-to-book ratio of $i$-th stock at the end of day $t_\ell$ from \eqref{def : rho}.
	
	\item $g_i(t_\ell), ~ h_i(t_\ell)$ : the decomposition of $\beta_i(t_\ell)$ in \eqref{eq : beta}. They are computed by recurrence relations $g_i(t_\ell) = g_i(t_{\ell-1}) + \big(\beta_i(t_\ell)-\beta_i(t_{\ell-1})\big)^+$ and $h_i(t_\ell) = h_i(t_{\ell-1}) + \big(\beta_i(t_\ell)-\beta_i(t_{\ell-1})\big)^-$ with initial values $g_i(t_0) = \beta_i(t_0)$, $h_i(t_0) = 0$.
	
	\item $\pi_i(t_\ell)$ : the portfolio weight of $i$-th stock at the end of day $t_\ell$. We first compute trading strategies \eqref{eq : varphi disconti} or \eqref{eq : psi disconti} for a generating function depending on the quantities listed 3-6 above and then calculate the portfolio weights via \eqref{def : portfolio}. We occasionally obtain the portfolio weights directly from the quantities in 3-6, as in \eqref{def : book value portfolio}.
	
	\item $W(t_\ell)$ : the total value of the portfolio at the end of day $t_\ell$. The transaction, or rebalancing, of our portfolio is made at the beginning of each day, so that the monetary value of the portfolio at the end of day $t_\ell$ can be calculated as
	\begin{equation*}
	W(t_\ell) = \sum_{i=1}^d W(t_{\ell-1})\pi_i(t_{\ell-1}) \frac{S_i(t_\ell)}{S_i(t_{\ell-1})}.
	\end{equation*}
	In order to compare the performance of our portfolios to the market portfolio, we normalize our initial wealth as $W(0) = \Sigma(0)$.
\end{enumerate}

When writing the trading strategy corresponding to the portfolio $\pi$ by $\varphi$, the relative value of \eqref{def : wealth process} represents the ratio between the monetary value of our portfolio and the total market capitalization:
\begin{equation*}
V^{\pi}(\cdot) = V^{\varphi}(\cdot)
= \sum_{i=1}^d \varphi_i(\cdot)\mu_i(\cdot) 
= \sum_{i=1}^d \varphi_i(\cdot)\frac{S_i(\cdot)}{\Sigma(\cdot)}
= \frac{W(\cdot)}{\Sigma(\cdot)}.
\end{equation*}

\medskip

\subsection{Book value portfolio}

\begin{figure}[H]
	\centering	
	\includegraphics[width=.7\linewidth]{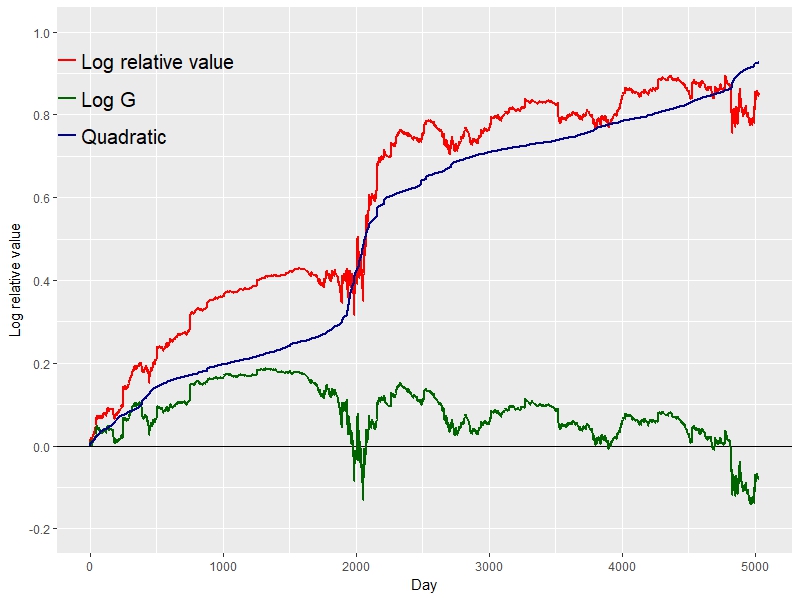} \label{fig: book value}
	\caption{Decomposition of the log relative value of the book value portfolio}
	\label{fig: BVP}
\end{figure}

We give the detailed empirical analysis of the book value portfolio of Example~\ref{ex: book value portfolio}. The red graph in Figure~\ref{fig: BVP} shows the log relative wealth $\log V^{\beta}$ on the left-hand side of \eqref{G in BVP}. The green and blue graphs represent the log of the generation function $G$ on the right-hand side of \eqref{G in BVP} and the drift process in \eqref{Quad in BVP}, respectively, in the decomposition of $\log V^{\beta}$. We note that the integral in \eqref{Beta term in BVP} vanishes in our dataset, as each integrator $\beta^c_i$ is a constant function.

Since the term \eqref{Quad in BVP} is derived from the quadratic variation term
\begin{equation*}
-\frac{1}{2} \int_0^t \sum_{i, j=1}^d \frac{D^2_{\mu_i, \mu_j}G(\mu, g, h)(s-)}{G(\mu, g, h)(s-)} d\langle \mu_i, \mu_j \rangle(s),
\end{equation*}
along with the fact that $G$ of \eqref{def : book value G} is concave in $\mu$, the blue graph has been increased over 20 years as expected.

\medskip

\subsection{Market-to-book ratio weighted portfolio}

We present the empirical results of the market-to-book ratio weighted portfolio $\Pi^{(p)}$ of \eqref{def : MTBR weighted portfolio}, along with the diversity weighted portfolio $\pi^{(p)}$ of \eqref{def : diversity weighted portfolio}, in Example~\ref{ex: market-to-book weighted portfolio}.

\begin{figure}[H]
	\centering	
	\subfloat[Market-to-book ratio weighted portfolio]
	{\includegraphics[width=.49\linewidth]{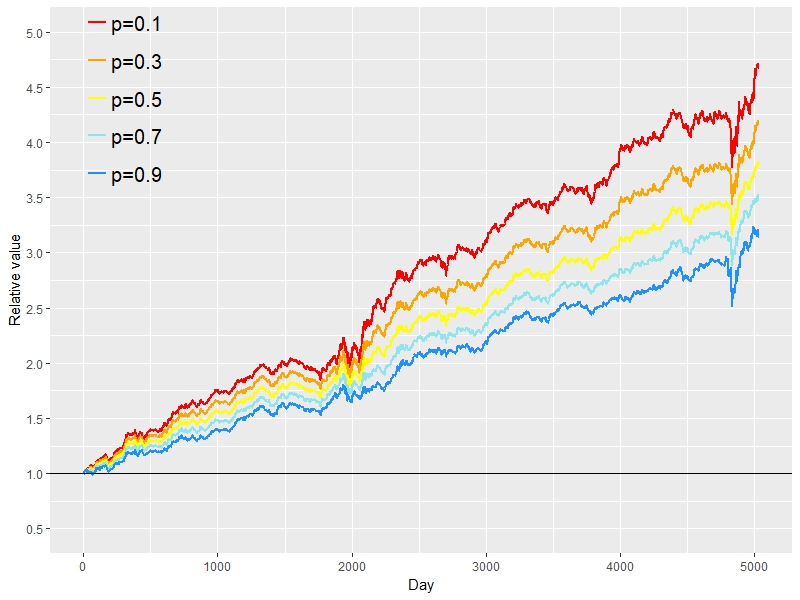}} \label{fig: MTB}
	\subfloat[Diversity-weighted portfolio]
	{\includegraphics[width=.49\linewidth]{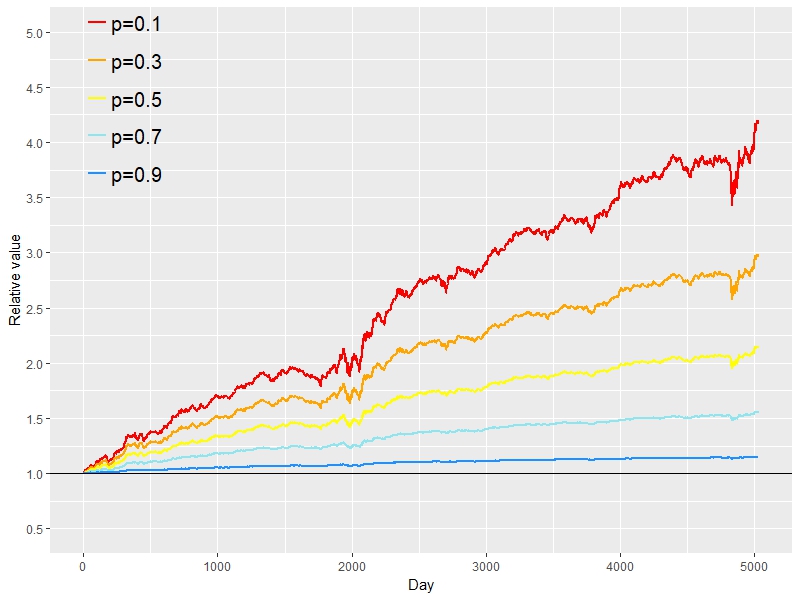}} \label{fig: diversity}\\
	\caption{Market-to-book ratio and diversity weighted portfolios with different $p$ values}
	\label{fig: weighted portfolios}
\end{figure}

Figure~\ref{fig: weighted portfolios} (a) shows the relative value processes of the market-to-book ratio weighted portfolios $\Pi^{(p)}$ with different values of $p$. The portfolio with a smaller value of $p$ tends to perform better. This clearly shows the ``value factor''; overweighting of the stocks with small market-to-book ratios~(and underweighting of the stocks with big market-to-book ratios) is preferable for the given data of $d=500$ stocks over 20 years.

\smallskip

In order to compare this value factor with ``size factor'', which is well-known in Stochastic Portfolio Theory, the relative values of the diversity-weighted portfolios $\pi^{(p)}$ with the same $p$ values are given in Figure~\ref{fig: weighted portfolios} (b). We note that both $\Pi^{(p)}$ and $\pi^{(p)}$ approach the equal-weighted portfolio as $p \downarrow 0$, and $\pi^{(1)}$ is just the market portfolio when $p = 1$.

\medskip

\medskip

\subsection{Portfolios depending on ranks of market-to-book ratios}

In this subsection, we compare the performances of portfolios described in Example~\ref{ex: EW}.

\smallskip

\begin{figure}[H]
	\centering	
	\subfloat[Equally-weighted portfolio of 250 stocks by ranks of $\rho$]
	{\includegraphics[width=.49\linewidth]{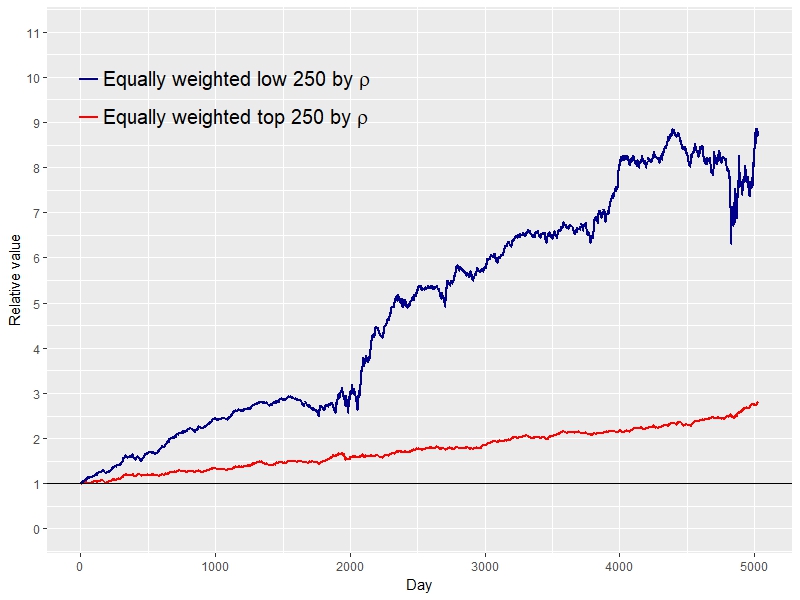}} \label{fig: rho250}
	\subfloat[Equally-weighted portfolio of 250 stocks by ranks of $\mu$]
	{\includegraphics[width=.49\linewidth]{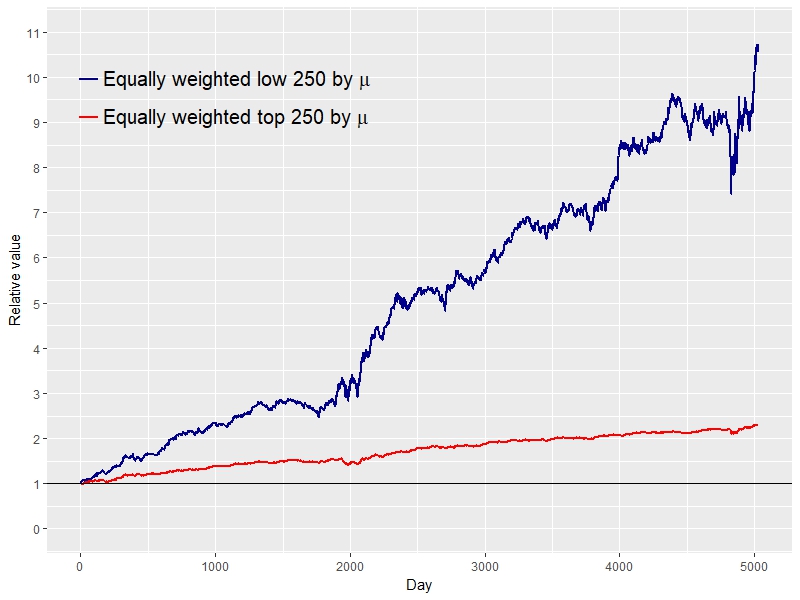}} \label{fig: mu250}\\
	\caption{Relative values of equally-weighted portfolios composed of the top and the bottom halves by ranks of market-to-book ratios and sizes}
	\label{fig: rank}
\end{figure}

We divide total of $d = 500$ stocks every trading day into two groups of $250$ stocks, according to the ranks of market-to-book ratios. The log relative values of two equally-weighted portfolios, generated from these groups, are shown in Figure~\ref{fig: rank} (a). Being as equally-weighted portfolios, both of them have been outperforming the market, but the return was higher for investment in the stocks with lower market-to-book ratios. This shows another evidence of the value factor in portfolio returns.

In order to compare this ``market-to-book ratio effect'' with the ``size effect'', we construct two more portfolios, equally-weighted for the top group and the bottom group, respectively, each composed of $250$ stocks but ranked by size. Let us recall the notations \eqref{def : ranked mu} and \eqref{def : permutation of mu} regarding the ranked relative capitalizations. Then, the graphs in Figure~\ref{fig: rank} (b) describe the relative values of the portfolios
\begin{equation*}
\hat{\pi}^{\psi}_i(t) = \sum_{k=1}^{250} \frac{I_{\{p_t(k)=i\}}}{250}, \qquad i = 1, \cdots, d,
\end{equation*}
and
\begin{equation*}
\underaccent{\hat}{\pi}^{\psi}_i(t) = \sum_{k=d-249}^{d} \frac{I_{\{p_t(k)=i\}}}{250}, \qquad i = 1, \cdots, d,
\end{equation*}
respectively.

\smallskip

The graphs in (a) and (b) of Figure~\ref{fig: rank} behave similar, but the gap between the performances of two equally-weighted portfolios generated from the top and bottom groups is a bit smaller when the ranks of the market-to-book ratios are used to separate the groups. This shows that the value factor has a milder effect than the size factor. The last relative values of these portfolios are summarized in the following table.

\begin{table}[H]
	\centering
	\begin{tabular}{|c|c|c|}
		\hline
		Equally-weighted portfolios & ~~~~~Top 250~~~~~ & ~~~~~Bottom 250~~~~~ \\ \hline
		By ranks of $\rho$         & 2.77319  & 8.74458     \\ \hline
		By ranks of $\mu$          & 2.28933 & 10.5868     \\ \hline
	\end{tabular}
	\caption{Relative values on the last day, $V(T)$ of portfolios in Figure~\ref{fig: rank}}
\end{table}

\bigskip

\subsection*{Acknowledgment}
The author is deeply grateful to Ioannis Karatzas and Nicola Doninelli for suggesting this topic and for numerous discussions regarding the material in this paper. The author is also indebted to Adrian Banner and Robert Fernholz for helpful comments, Johannes Ruf for detailed reading and suggestions which improved this paper.

\bigskip

\bibliography{aa_bib}
\bibliographystyle{apalike}

\end{document}